\documentclass[iop,apj]{emulateapj}

\usepackage{epsfig}
\usepackage{graphics,graphicx}
\usepackage{multirow}
\usepackage{gensymb}
\usepackage{amsmath,amssymb}
\usepackage{bm}
\usepackage{natbib}
\usepackage[outdir=./]{epstopdf}
\usepackage[breaklinks,colorlinks,urlcolor=blue,citecolor=blue,linkcolor=blue]{hyperref}
\usepackage{chngcntr}

\def\h1{H{\textsc i}}
\def\mstar{$M_{\mathrm{star}}$}
\def\mhi{$M_{\mathrm{H{\textsc i}}}$}
\def\mhic{$M_{\mathrm{H{\textsc i}},\,\mathrm{c}}$}
\def\sfehi{SFE$_{\mathrm{H{\textsc i}}}$}
\def\fhi{$f_{\mathrm{H{\textsc i}}}$}
\def\mh2{$M_{\mathrm{H_2}}$}
\def\mgas{$M_{\mathrm{gas}}$}
\def\mdust{$M_{\mathrm{dust}}$}
\def\msun{$M_{\odot}$}

\shorttitle{HI Observation of Major Merger Pairs at z = 0}
\shortauthors{P.~Zuo et al.}

\begin{document}


\title{H$\textsc I$ Observations of Major-merger Pairs at z = 0: atomic gas and star formation}


\author{Pei Zuo\altaffilmark{1,4}, Cong K. Xu\altaffilmark{2,1}, Min S.~Yun\altaffilmark{3}, Ute Lisenfeld\altaffilmark{5}, Di Li\altaffilmark{1,4,6}, and Chen Cao\altaffilmark{7, 8}}

\affil{
$^1$ National Astronomical Observatories, Chinese Academy of Sciences, Beijing 100101, China\\
$^2$ Chinese Academy of Sciences South America Center for Astronomy, China-Chile Joint Center for Astronomy, Camino El Observatorio 1515, Las Condes, Santiago, Chile\\
$^3$ Department of Astronomy, University of Massachusetts, Amherst, MA 01003, USA\\
$^4$ University of Chinese Academy of Sciences, Beijing 100049, China\\
$^5$ Departamento de F\'{i}sica Te\'{o}rica y del Cosmos, Universidad de Granada, Spain and Instituto Carlos I de F\'{i}sica T\'{e}orica y
Computacional, Facultad de Ciencias, 18071 Granada, Spain \\
$^6$ CAS Key Laboratory of FAST, NAOC, Chinese Academy of Sciences, Beijing,100101, China  \\
$^7$ School of Space Science and Physics, Shandong University, Weihai, Shandong 264209, China \\
$^8$ Shandong Provincial Key Laboratory of Optical Astronomy \& Solar-Terrestrial Environment, Weihai, Shandong 264209, China \\
}

\email{Email: peizuo@nao.cas.cn, congxu@nao.cas.cn}


\altaffiltext{1}{National Astronomical Observatories, Chinese Academy of Sciences, Beijing 100101, China}
\altaffiltext{2}{South American Center for Astronomy, CAS, Camino El Observatorio 1515, Las Condes, Santiago, Chile}
\altaffiltext{3}{Department of Astronomy, University of Massachusetts, Amherst, MA 01003, USA}
\altaffiltext{4}{University of Chinese Academy of Sciences, Beijing 100049, China}
\altaffiltext{5}{Departamento de F\'{i}sica Te\'{o}rica y del Cosmos, Universidad de Granada, Spain and Instituto Carlos I de F\'{i}sica T\'{e}orica y
Computacional, Facultad de Ciencias, 18071 Granada, Spain}
\altaffiltext{6}{CAS Key Laboratory of FAST, NAOC, Chinese Academy of Sciences, Beijing,100101, China  }
\altaffiltext{7}{School of Space Science and Physics, Shandong University, Weihai, Shandong 264209, China }
\altaffiltext{8}{Shandong Provincial Key Laboratory of Optical Astronomy \& Solar-Terrestrial Environment, Weihai, Shandong 264209, China}

\hyphenpenalty=5000
\tolerance=1000

\begin{abstract}

    We present a study of the \h1\ gas content of a large
    K-band selected sample of 88 close major-merger pairs of galaxies
    (H-KPAIR) which were observed by $\it Herschel$. We obtained the
    21 cm \h1\ fine-structure emission line data for a total of 70
    pairs from this sample, by observing 58 pairs using the Green Bank
    Telescope (GBT) and retrieving the \h1\ data for an addition 12
    pairs from the literature.  In this \h1\ sample, 34 pairs are
    spiral-spiral (S+S) pairs, and 36 are spiral-elliptical (S+E).
    Based on these data, we studied the \h1-to-stellar mass ratio, the
    \h1\ gas fraction and the \h1\ star formation efficiency (\sfehi =
    star formation rate/\mhi) and searched for differences between S+S
    and S+E pairs, as well as between pairs with and without signs for
    merger/interaction.  Our results showed that the mean
    \h1-to-stellar mass ratio of spirals in these pairs is $=7.6\pm
    1.0 \%$, consistent with the average \h1 gas fraction of spiral
    galaxies in general.  The differences in the \h1\ gas fraction
    between spirals in S+S and in S+E pairs, and between spirals in
    pairs with and without signs of merger/interaction are
    insignificant ($< 1 \sigma$).  On the other hand, the mean \sfehi\
    of S+S pairs is $\sim4.6\times$ higher than that of S+E pairs.
    This difference is very significant ($\sim 4\sigma$) and is the
    main result of our study.  There is no significant difference in
    the mean \sfehi\ between galaxies with and without signs of
    merger/interaction.  The mean \sfehi\ of the whole pair sample is
    $10^{-9.55\pm 0.09}\ \mathrm{yr}^{-1}$,
    corresponding to a \h1\ consumption time of $3.5\pm0.7$~Gyrs.  

\end{abstract}





\section{INTRODUCTION}

It has been well documented that galaxy-galaxy interaction can induce enhanced star formation \citep{Larson1978, Keel1985, Kennicutt1987, Bushouse1988, Telesco1988, Sulentic1989, Xu1991,Barton2000, Lambas2003, Alonso2004, Nikolic2004, Li2008, Ellison2008, Xu2010}. 
More recently, \citet{Scudder2012} and \citet{Patton2013} found that wide galaxy pairs with separation as large as $\sim$80~kpc still show significant star formation rate (SFR) enhancement at $\sim$40\% level.
Early studies \citep{Hummel1981, Haynes1988, Bergvall2003} that failed to detect SFR enhancement in interacting galaxies may have suffered from biases in selecting the interacting galaxy sample and the control sample \citep[c.f.]{Xu2010, Ellison2010}.

Of particular interest are close (separation $\leq 20~h^{-1}~\mathrm{kpc}$) major mergers of galaxies of nearly equal mass (primary-to-secondary mass ratio $\lesssim 3$). 
Most extreme starbursts such as ultra-luminous infrared galaxies (ULIRGs) are close major mergers \citep{Sanders1996, Dasyra2006}. For a sample of K-band selected close major-merger pairs, 
IR observations carried out using $\it Spitzer$ \citep{Xu2010} and $\it Herschel$ \citep{Cao2016} found that in these pairs the average specific SFR (sSFR = SFR/\mstar) in spirals is a factor of $\gtrsim$2 higher than that of their counterparts in the control sample. 
Furthermore, spirals in spiral-spiral pairs (S+S pairs) are strongly enhanced with the mean sSFR $\gtrsim$3 times higher than that of control galaxies, 
but spirals in mixed spiral-elliptical pairs (S+E pairs) do not show any significant SFR enhancement compared to the control galaxies. 
Using $\it WISE$ and $\it Herschel$ data, \citet{Domingue2016} also found 
that spirals in S+S pairs exhibit
significant enhancements in interstellar radiation field and dust
temperature while spirals in S+E pairs do not.

Why is the sSFR enhancement of spirals in mixed S+E pairs different from that of spirals in S+S pairs? 
If the enhancement is purely due to gravitational tidal effect, 
then the spirals in S+E pairs should behave similarly to spirals in S+S pairs unless the former have systematically less cold gas (i.e.\ the fuel for star formation) than the latter.
This possibility has been tested by \citet{Cao2016}. 
Using fluxes in six $\it Herschel$ bands (70, 100, 160, 250, 350, and 500 $\mu m$), 
they estimated dust mass (\mdust) and, assuming a constant dust-to-gas mass ratio, total gas mass (\mgas) for the paired spirals. 
They found only marginal evidence for spirals in S+E pairs having slightly lower gas content than those in S+S pairs ($\delta \mathrm{log}(M_\mathrm{gas}/M_\mathrm{star}) = -0.14 \pm 0.10$). 
It appears that the difference between the sSFR enhancements of spirals in mixed S+E pairs and of those in S+S pairs is mainly due to their different star formation efficiency (SFE = SFR/\mgas), 
suggesting significant roles for non-tidal effects (e.g.\ collision between gas in two galaxies) in the interaction induced star formation.

In this paper, we represent a study on the \h1\ gas content of pairs. 
The main science goal is to constrain the relation between \h1\ gas content and SFR enhancement, 
and check the consistency with the relation between gas content (estimated using the dust mass) and SFR enhancement obtained in the $\it Herschel$ study \citep{Cao2016}. 
In section 2, 3 and 4, we describe the sample, GBT observations, and data reduction. 
Literature data are described in Section 5. 
Main results are presented in Section 6. 
Section 7 is devoted to discussions. 
Section 8 is the summary. 
Through out this paper, we adopt the $\Lambda$-cosmology with $\Omega_m = 0.3$ and $\Omega_\Lambda = 0.7$, and H$_0 = 70$ (km~s$^{-1}$~Mpc$^{-1}$).

\section{THE H-KPAIR SAMPLE}

The KPAIR is an unbiased and large sample of 170 close
major-merger galaxy pairs selected in the K-band, from cross matches
between the Two Micron All Sky Survey (2MASS) and the Sloan Digital
Sky Survey (SDSS)-DR5 galaxies \citep{Domingue2009}.  The parent
sample includes 77,451 galaxies of Ks $\leq 13.5$ mag, with a sky
coverage of 5800 deg$^2$ and redshift completeness of 86\%.  The
selection criteria are: (1) Ks magnitude of the primary is not
fainter than 12.5;  (2) at least one component has a measured
redshift;  (3) if both components have measured redshifts, the velocity
difference is not larger than 1000 $\rm km\; s^{-1}$;  (4) the Ks difference
between the two galaxies is not larger than 1 mag; (5) the
projected separation is in the range of $\rm 5\; h^{-1}\; kpc \leq r
\leq 20\; h^{-1}\; kpc$.  When only one component has a measured
redshift, the separation is calculated according to that redshift and
the angular separation of the components.  Visual
inspections, complemented by results of an automatic algorithm,
classified 62 pairs as S+S, 56 as S+E, and 52 as E+E.

The H-KPAIR sample \citep{Cao2016}
includes all S+S and S+E pairs in the original KPAIR sample
that have (1) measured redshifts for both components, 
(2) relative velocity $\rm < 500\; km\; s^{-1}$, and (3)
pair recession velocity $\rm < 2000\; km\; s^{-1}$.
It contains 88 pairs (44 S+S and 44 S+E).
We did some of these comparison in \citet{Xu2010} and \citet{Cao2016}. 
We will do a comprehensive comparison between S+S and S+E in a future paper.

\section{OBSERVATIONS}

For 67 pairs, the 21~cm \h1\ fine-structure line observations were carried out using the National Radio Astronomy Observatory (NRAO)\footnote{The National Radio Astronomy Observatory is a facility of the National Science Foundation operated under cooperative agreement by Associated Universities, Inc.} Robert C. Byrd Green Bank 110-m Telescope (GBT)\footnote{The Green Bank Observatory is a facility of the National Science Foundation operated under cooperative agreement by Associated Universities, Inc.} Spectrometer in the L-band (1.15-1.73 GHz) between 2012 August and 2013 January. 
For each object, the data were collected in $\sim$2~hrs on-off source pairs with 12.5~MHz bandwidth. 
The two spectral windows were centered at the same frequency (1420.4058~MHz). 
Using 9-level sampling and two IFs, the observations provide 1.5~kHz (0.3~km~s$^{-1}$) spectral resolution for the dual polarization L-band system. 
The beam size is $9\arcmin \times 9\arcmin$. 
GBT has a well-calibrated structure and a stable gain at the 21~cm wavelength. 
We observed 3C 286 as the primary flux calibrator to monitor the instrumental performance. 
This observation of a bright calibration source verified the stability of the telescope gain factor. 
As a test, we also observed 5 nearby normal galaxies: 
NGC895, NGC2718, NGC3027, UGC10014, and NGC6140. 
Comparisons with literature show a systematic difference of $\sim$15\% between our measurements and data in the literature (see Appendix A), suggesting a minor deviation in the calibration. 
This shall not affect our main conclusions significantly.

\section{DATA REDUCTION}

\begin{figure*}
\includegraphics[width = 1.0\textwidth]{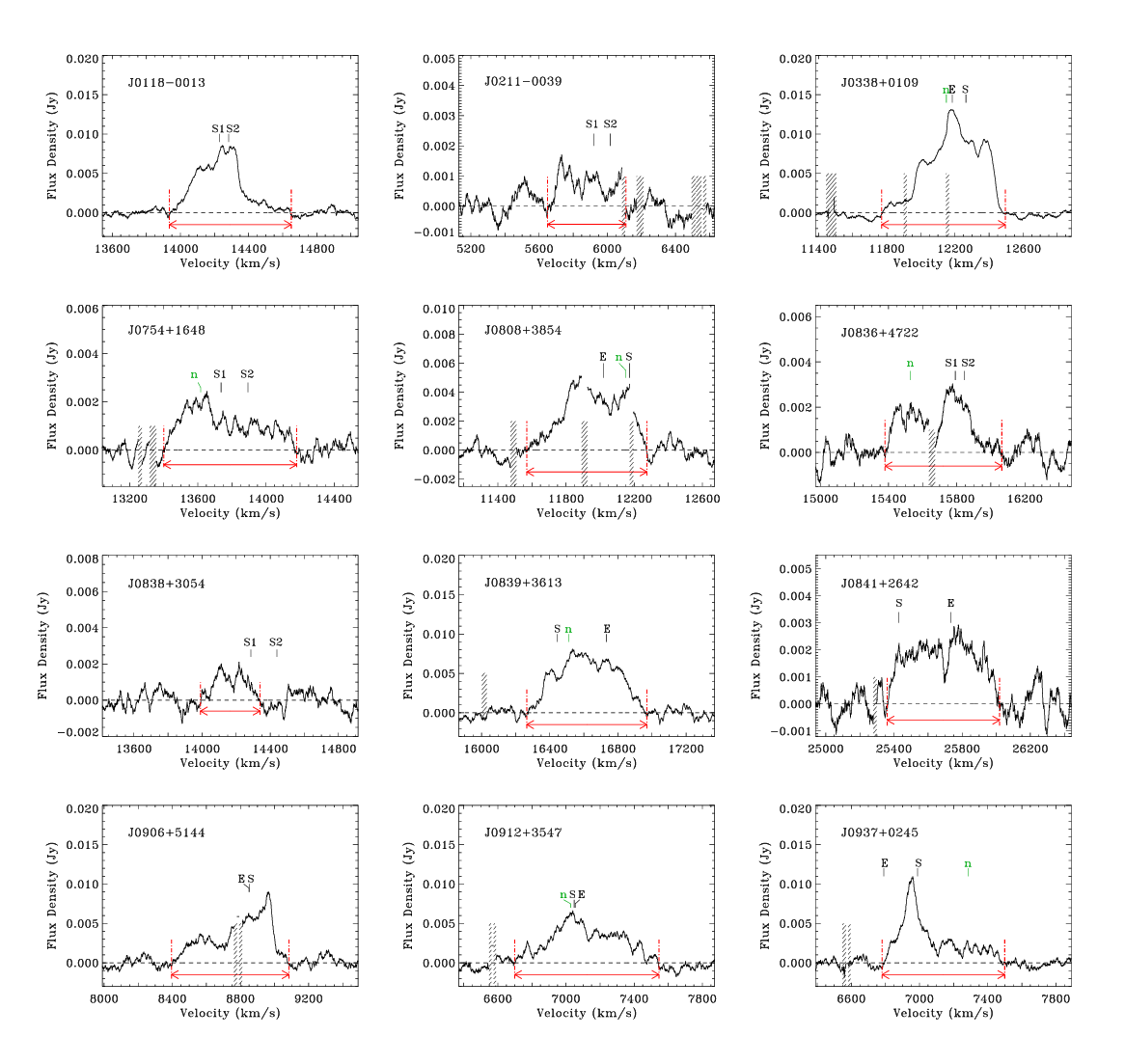}
\caption{\h1 profiles of major-merger pairs observed with 100m GBT. Optical redshifts of individual galaxies are marked with short vertical lines on the observed spectra. For S+E pairs, ``S" and ``E" are for the S and E components, respectively. For the S+S pairs, ``S1" represents the western galaxy and ``S2" represents the eastern one. Short vertical lines in green denoted by n, n1, n2, ..., mark the optical redshifts of neighboring galaxies in the beam. The regions with hash marks are frequencies where the data are affected by RFI spikes. The region between the two red dot-dash lines represents the range of the intensity flux integration.
(The complete figure set (5 plots) is available.)}
\label{fig:spectra_hi}
\end{figure*}

The \h1 spectra were reduced using GBTIDL \citep{Marganian2006}. 
We used the Jy/K calibration to convert the \h1\ line fluxes to the units of Jy, applying an atmospheric opacity of 0.008 and aperture efficiency of 0.71. 
The scans and channels with Radio Frequency Interferences (RFIs) were flagged. 
For each polarization, the data were accumulated and averaged together. 
A polynomial of the order of 3~-~8 was used to fit the baseline over a range of $\approx4500$ channels for every pair. 
The Hanning-smoothed and then decimated spectra were used to subtract the baseline. 
The velocity resolution is $\sim$30~km~s$^{-1}$ per channel after boxcar smoothing.
The two polarizations were then averaged together to produce the full intensity spectra shown in Figure~\ref{fig:spectra_hi}.

Some observations were significantly affected by RFIs. 
For 9 pairs (J0913+4742,  J0926+0447, J1010+5440, J1020+4831, 
J1137+4728, J1148+3547, J1205+0135, J1505+3427, J1628+4109),
the RFIs are so severe that no informative signals could be
extracted from the data. 
The pairs are excluded from our analysis. 
Forty-six targets are detected with the \h1\ 21~cm emission peak $>3\sigma$. 
Their \h1 masses were calculated using the relationship of $M_{\mathrm{\h1}}=2.36\times10^{5}D^2(S\Delta V)M_{\odot}$ \citep{Condon2016}, 
where $D$ is the luminosity distance in Mpc and $S\Delta V$ is the velocity integrated \h1\ flux density in Jy km s$^{-1}$. 
The \h1 velocity range was visually decided, 
with the constraint that the center is within $\pm400$~km~s$^{-1}$ from the system velocity (optical) of the target. 
The \h1 masses of the 12 undetected pairs were calculated based on the 3$\sigma$ upper limits of the spectral line with assumed line-width of $\sim$800~km~s$^{-1}$. 
The error was estimated by the quadratic sum of the 10\% systematic uncertainty (mostly due to the calibration and the baseline subtraction error) and the measured rms noise. 
We also measured the mean \h1 velocity (weighted by signal) and the $W20$ of the spectral line for detected pairs.

For the 46 detected pairs, 
we visually inspected their SDSS images and found that 22 have neighboring spiral galaxies
with redshifts inside the bandwidth of the \h1\ observation and locations 
 within 10\arcmin\ from the pair center (Figure~\ref{fig:appb_sdss} in Appendix B). 
We performed our search down to the 
limit of 17.71 mag ($i-$band), 18.35 mag ($g-$band) and 17.78 mag ($r-$band). 
All spiral galaxies in the search radius and brighter than these limits had observed redshifts. 
As the beam of GBT is 9\arcmin, 
contamination due to blending could be significant for these sources, 
therefore a correction was carried out based on the algorithm developed by
\citet{Zhang2009}. Details about the contamination 
correction are presented in Appendix B.
Most of the neighboring galaxies cause minor corrections. 
The average and the range of the factor by which \mhi\ 
was changed due to the correction are 1.26 and 1.03-1.84, 
respectively. Also, as a test for the algorithm, 
we found that for paired galaxies 
in our sample the ratio between estimated and observed
\h1\ mass is consistent with being unity (0.8$\pm$0.2). 
The \h1\ mass after this correction was listed as \mhic\ in column (7) in Table~\ref{tbl:gbt_obs}.

\section{LITERATURE DATA}

The 21 pairs in H-KPAIR that we did not observe with GBT 
were covered by previous \h1\ emission line observations.
However, detailed inspections showed that among them 9 pairs 
(J0915+4419, J1015+0657, J1150+1444, J1211+4039, J1219+1201,
 J1429+3534, J1506+0346, J1514+0403, J1608+2529)
are in galaxy groups and the \h1\ observations were 
not pointed to the pairs but to neighboring galaxies in the same group, as shown in Figure~\ref{fig:appb_sdss}. 
Therefore their \h1\ mass
are too uncertain and they are excluded from our analysis.
For each of the remaining 12 pairs, the \h1\ data 
collected from the literature are listed in Table~\ref{tbl:lit_obs}. 
No neighboring spiral galaxies that can cause significant \h1 contaminations 
were found for these pairs (Figure~\ref{fig:appb_sdss} in Appendix B).

\newcommand{\noprint}[1]{}
\newcommand{\figsetstart}{{\bf Fig. Set} }
\newcommand{\figsetend}{}
\newcommand{\figsetgrpstart}{}
\newcommand{\figsetgrpend}{}
\newcommand{\figsetnum}[1]{{\bf #1.}}
\newcommand{\figsettitle}[1]{ {\bf #1} }
\newcommand{\figsetgrpnum}[1]{\noprint{#1}}
\newcommand{\figsetgrptitle}[1]{\noprint{#1}}
\newcommand{\figsetplot}[1]{\noprint{#1}}
\newcommand{\figsetgrpnote}[1]{\noprint{#1}}
 


\figsetgrpstart
\figsetgrpnum{1.1}
\figsetgrptitle{Image of M1}
\figsetplot{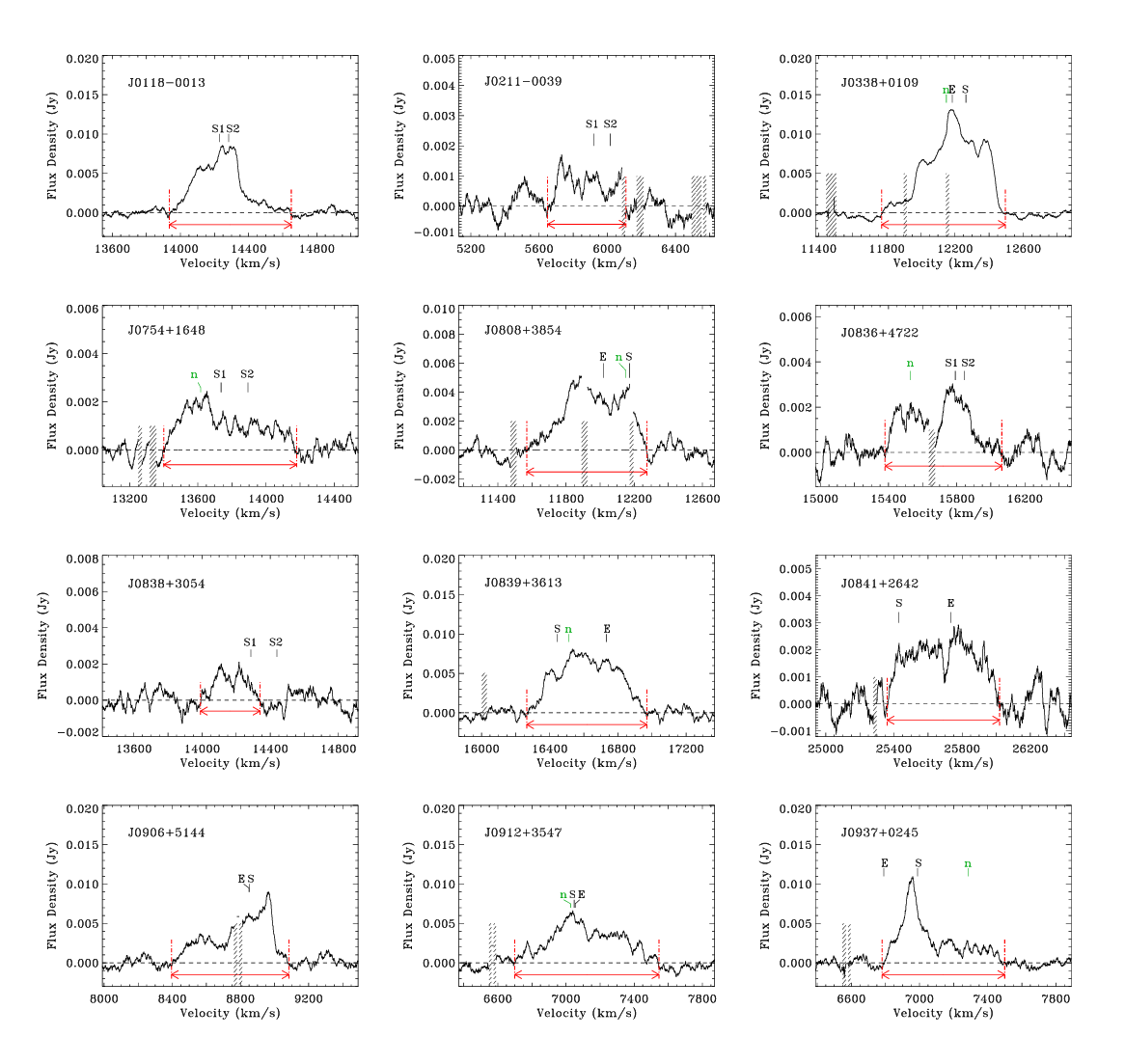}
\figsetgrpnote{Images collected by pzuo.}
\figsetgrpend

\figsetgrpstart
\figsetgrpnum{1.2}
\figsetgrptitle{Image of M2}
\figsetplot{f1_2.ps}
\figsetgrpnote{Images collected by pzuo.}
\figsetgrpend

\figsetgrpstart
\figsetgrpnum{1.3}
\figsetgrptitle{Image of M3}
\figsetplot{f1_3.ps}
\figsetgrpnote{Images collected by pzuo.}
\figsetgrpend

\figsetgrpstart
\figsetgrpnum{1.4}
\figsetgrptitle{Image of M4}
\figsetplot{f1_4.ps}
\figsetgrpnote{Images collected by pzuo.}
\figsetgrpend

\figsetgrpstart
\figsetgrpnum{1.5}
\figsetgrptitle{Image of M5}
\figsetplot{f1_5.ps}
\figsetgrpnote{Images collected by pzuo.}
\figsetgrpend

\figsetend

\begin{table*}
\tiny
\begin{center}
\caption{Pairs in GBT observations.\label{tbl:gbt_obs}}
\begin{tabular}{crrrrrrrrrrrrrr}
\tableline\tableline
(1) & (2) & (3) & (4) & (5) & (6) & (7) & (8) & (9) & (10)  & (11)  & (12) \\
Pair ID & R.A. & decl. & $V_{\mathrm{optical}}$  & $V_{\mathrm{H{\textsc i}}}$  & \mhi\  & $M_{\mathrm{H{\textsc i},\,c}}$  &  $W20$  & 100$\times$\mdust\  & 
 \mstar\   &    SFR    &     
 Type \\
(H-KPAIR)  & (J2000)  & (J2000)  & (km\,s$^{-1}$)  & (km\,s$^{-1}$)  & (10$^9${\it{M$_{\odot}$}})  & (10$^9${\it{M$_{\odot}$}})  & (km\,s$^{-1}$)& (10$^9${\it{M$_{\odot}$}})  & (10$^9${\it{M$_{\odot}$}})  & ({\it{M$_{\odot}$}} yr$^{-1}$)  &
   \\

\tableline

J0118-0013 & 01:18:34.9 & -00:13:50 & 14160 &    14235 & 22.64 $\pm$ 2.29 &  &  389.38 & 26.51 &  140.17 & 61.83 & SS \\
J0211-0039 & 02:11:07.4 & -00:39:17 & 5970 &     5874 &  0.45 $\pm$ 0.06 &  & *301.24 &  4.54 &  76.05 & 2.95 & SS \\
J0338+0109 & 03:38:12.6 & +01:09:55 & 12240 &    12194 & 29.76 $\pm$ 2.98 &  21.72 & 491.08 & 13.49 &  50.12 &  5.02 & SE \\
J0754+1648 & 07:54:32.1 & +16:48:28 & 13812 &    13741 &  8.42 $\pm$ 0.90 &  7.24 & *612.13 & 42.90 & 232.46 & 30.02 & SS \\
J0808+3854 & 08:08:34.7 & +38:54:52 & 12040 &    11964 &  12.21 $\pm$ 1.26 &  11.21 & 519.60 &  2.29 &  51.29 &  1.35 & SE \\
J0836+4722 & 08:36:45.4 & +47:22:14 & 15768 &    15684 &  11.03 $\pm$ 1.28 &  6.00 & 507.72 & <4.95 & 230.40 & 0.82 & SS \\
J0838+3054 & 08:38:17.8 & +30:54:57 & 14344 &    14167 &  2.52 $\pm$ 0.54 &  & 294.54 &  4.26 &  162.11 &  4.21 & SS \\
J0839+3613 & 08:39:00.5 & +36:13:10 & 16569 &    16619 & 46.29 $\pm$ 4.70 &  44.05 & 551.70 & 13.49 &  81.28 &  1.91 & SE \\
J0841+2642 & 08:41:50.1 & +26:42:52 & 25600 &    25685 &  8.46 $\pm$ 2.10 &  & 617.95 &  4.90 & 257.04 &  1.43 & SE \\
J0906+5144 & 09:06:03.9 & +51:44:24 & 8737 &     8795 &  8.30 $\pm$ 0.87 &  & 538.90 &  6.92 &  39.81 &  1.26 & SE \\
J0912+3547 & 09:12:36.6 & +35:47:32 & 7056 &     7112 &  5.98 $\pm$ 0.62 &  5.61 & 629.60 & <0.37 &  17.38 & <0.04 & SE \\
J0937+0245 & 09:37:44.6 & +02:45:14 & 6890 &     7056 &  5.23 $\pm$ 0.55 &  5.01 &  *275.36 & 23.44 & 144.54 &  9.83 & SE \\
J1022+3446 & 10:22:56.5 & +34:46:51 & 16761 &    16733 & 15.25 $\pm$ 1.75 &  & 572.21 & 7.59 &  143.43 & 5.65 & SS \\
J1023+4220 & 10:23:36.7 & +42:20:55 & 13659 &    13651 & 32.74 $\pm$ 3.33 & 28.44 & *451.30 & 15.11 &  100.05 & 13.36 & SS \\
J1027+0114 & 10:27:29.6 & +01:15:02 & 6670 &     6718 &  6.01 $\pm$ 0.62 &  5.63 &  *139.39 &  2.75 &  26.30 &  1.91 & SE \\
J1032+5306 & 10:32:53.2 & +53:06:50 & 19186 & ... & <3.06 &  & ... & <2.85 &  104.71 &  1.03 & SE \\
J1033+4404 & 10:33:30.7 & +44:04:27 & 15658 &    15791 &  24.21 $\pm$ 2.62 &  & *432.10 & 34.32 & 216.94 & 18.53 & SS \\
J1036+5447 & 10:36:43.4 & +54:47:42 & 13743 & ... & <2.39 &  & ... & <1.93 &  60.26 & <0.17 & SE \\
J1039+3904 & 10:39:24.3 & +39:04:53 & 13017 & ... & <1.32 &  & ... &  1.93 &  52.48 & <0.17 & SE \\
J1045+3910 & 10:45:24.9 & +39:10:09 & 7879 & ... & <0.72 &  & ... &  4.27 &  42.66 &  0.67 & SE \\
J1051+5101 & 10:51:44.1 & +51:01:25 & 7325 & ... & <0.57 &  & ... & <0.45 &  63.09 & <0.16 & SE \\
J1059+0857 & 10:59:58.9 & +08:57:28 & 18490 & ... & <3.20 &  &  ... & <2.98 &  72.44 & <0.32 & SE \\
J1101+5720 & 11:01:43.6 & +57:20:19 & 14208 &    14130 &  7.90 $\pm$ 1.13 &  & 280.37 & <2.34 &  34.67 & <2.01 & SE \\
J1106+4751 & 11:06:50.1 & +47:51:10 & 19464 &    19516 & 47.27 $\pm$ 5.01 &  & 647.25 & 16.60 & 228.83 & 8.47 & SS \\
J1120+0028 & 11:20:47.3 & +00:28:10 & 7295 &     7308 &  7.81 $\pm$ 0.79 &  5.59 & 449.42 &  1.16 &  94.00 & 1.12 & SS \\
J1125+0226 & 11:25:17.1 & +02:26:54 & 14730 &    14713 & 20.91 $\pm$ 2.23 &  & 613.87 & 10.94 &  145.81 & 1.13 & SS \\
J1127+3604 & 11:27:33.8 & +36:04:01 & 10528 &    10397 & 14.87 $\pm$ 1.51 &  11.97 & 555.50 & 21.55 & 224.07 &  5.76 & SS \\
J1144+3332 & 11:44:03.8 & +33:32:20 & 9489 &     9519 &  4.64 $\pm$ 0.54 &  2.77 & 336.63 &  3.80 &  22.39 &  1.11 & SE \\
J1150+3746 & 11:50:13.7 & +37:46:20 & 16587 &    16491 & 39.70 $\pm$ 4.03 &  & 923.50 & 13.72 &  158.31 &  4.10 & SS \\
J1154+4932 & 11:54:23.0 & +49:32:48 & 21200 & ... & <4.41 &  & ... &  3.11 &  89.13 & <0.42 & SE \\
J1202+5342 & 12:02:04.8 & +53:42:40 & 19290 &    19269 &  12.57 $\pm$ 1.64 &  &  *282.80 & 10.72 &  75.86 &  1.78 & SE \\
J1243+4405 & 12:43:39.1 & +44:05:52 & 12412 &    12455 &  5.77 $\pm$ 0.67 &  & *255.70 &  4.36 &  69.18 &  1.17 & SE \\
J1252+4645 & 12:52:51.1 & +46:45:28 & 18346 &    18246 & 5.80 $\pm$ 1.41 &  & *111.90 & 11.75 & 112.20 &  1.84 & SE \\
J1301+4803 & 13:01:17.5 & +48:03:33 & 9018 &     8938 &  8.75 $\pm$ 0.89 &  &  361.96 &  6.86 &  52.78 &  8.57 & SS \\
J1313+3910 & 13:13:14.5 & +39:10:37 & 21475 & ... & <4.70 &  & ... & <3.01 &  83.18 & <0.68 & SE \\
J1315+4424 & 13:15:15.6 & +44:24:26 & 10740 &    10634 &  5.58 $\pm$ 0.63 &  & 352.93 & 10.16 &  163.35 & 10.63 & SS \\
J1332-0301 & 13:32:55.9 & -03:01:37 & 14643 &    14145 & 14.15 $\pm$ 1.89 & 10.18 & *165.80 & 12.81 & 122.97 & 11.45 & SS \\
J1346-0325 & 13:46:21.1 & -03:25:23 & 7030 &     6883 &  4.40 $\pm$ 0.46 &  2.57 &  *109.61 &  2.82 &  48.98 &  0.28 & SE \\
J1400-0254 & 14:00:37.3 & -02:54:27 & 7390 &     7491 & 8.12 $\pm$ 0.87 &  & 669.25 & <1.09 &  75.29 & 0.08 & SS \\
J1400+4251 & 14:00:58.3 & +42:51:01 & 9939 &     9739 &  7.37 $\pm$ 0.79 &  & *418.83 & 17.48 &  91.10 & 17.99 & SS \\
J1405+6542 & 14:05:52.1 & +65:42:43 & 9208 &     9224 &  5.45 $\pm$ 0.60 &  3.31 & *304.06 &  6.31 &  19.95 &  0.73 & SE \\
J1407-0234 & 14:07:07.1 & -02:34:45 & 17020 &    16954 & 40.54 $\pm$ 4.29 &  & 659.52  & 10.23 &  95.50 &  1.49 & SE \\
J1424-0304 & 14:24:58.7 & -03:04:00 & 15400 &    15517 & 46.18 $\pm$ 4.73 &  & 702.46 & 17.36 & 213.11 & 4.07 & SS \\
J1433+4004 & 14:33:47.6 & +40:05:15 & 7871 &     7773 &  9.28 $\pm$ 0.95 & 8.67 & *269.16 & 19.34 &  148.76 &  9.67 & SS \\
J1500+4317 & 15:00:24.4 & +43:17:04 & 9399 &     9046 & 14.65 $\pm$ 1.50 & 14.03 & 661.15 & <0.90 &  53.70 & <0.10 & SE \\
J1510+5810 & 15:10:16.8 & +58:10:39 & 9303 & ... & <0.81 &  & ... & 4.79 &  81.74 & 2.04 & SS \\
J1523+3748 & 15:23:38.3 & +37:48:44 & 7048 & ... & <0.47 &  & ... &  0.89 &  13.80 &  0.47 & SE \\
J1526+5915 & 15:26:48.3 & +59:15:47 & 13525 & ... & <1.96 &  & ... & <1.16 &  58.88 & <0.14 & SE \\
J1528+4255 & 15:28:14.7 & +42:56:13 & 5530 &     5473 &  8.54 $\pm$ 0.86 & 7.21 & *490.79 &  8.42 &  141.32 &  3.07 & SS \\
J1552+4620 & 15:52:33.3 & +46:20:20 & 18060 &    17708 &  17.55 $\pm$ 1.93 &  & 554.70 & 14.45 &  83.18 &  3.53 & SE \\
J1556+4757 & 15:56:24.7 & +47:57:23 & 5850 &     5815 &  3.36 $\pm$ 0.35 &  2.89 & 831.43 &  2.34 &  14.12 &  1.50 & SE \\
J1558+3227 & 15:58:37.7 & +32:27:42 & 14679 &    14432 &  7.03 $\pm$ 0.85 &  &  342.53 & 10.89 &  113.93 & 4.81 & SS \\
J1602+4111 & 16:02:43.7 & +41:11:54 & 10026 &    10024 & 28.16 $\pm$ 2.84 &  27.45 & *490.18 & 18.62 &  94.76 & 13.50 & SS \\
J1614+3711 & 16:14:54.2 & +37:11:10 & 17450 &    17541 & 34.27 $\pm$ 3.65 & 31.37 & 723.05 & <1.93 & 131.83 &  0.46 & SE \\
J1635+2630 & 16:35:43.3 & +26:30:49 & 21207 &    21191 & 60.97 $\pm$ 6.30 &  53.61 & *342.6 & <3.33 & 169.82 &  0.90 & SE \\
J1637+4650 & 16:37:26.7 & +46:50:10 & 17915 &    17398 &  16.29 $\pm$ 1.80 &  & *532.28 & 31.34 & 277.47 &  3.73 & SS \\
J1702+1859 & 17:02:03.5 & +18:59:55 & 16965 & ... & <2.61 &  &  ... & <1.71 &  46.77 & <0.31 & SE \\
J1704+3448 & 17:04:50.9 & +34:48:57 & 17028 &    16990 & 15.43 $\pm$ 1.71 &  & *348.13 & 23.80 &  152.68 & 32.12 & SS \\

\tableline
\end{tabular}
\tablecomments{Descriptions of Columns: 
(1) Pair ID. The designations are ``H-KPAIR J0118-0013", etc. 
(2) R.A. (h:m:s, J2000). 
(3) decl. (d:m:s, J2000). 
(4) Optical velocity taken from SDSS or other telescopes. 
(5) \h1 mean velocity.
(6) \h1\ mass (10$^9${\it{M$_{\odot}$}}). 
(7) \h1\ mass after the correction of contamination due to
    neighboring galaxies (10$^9${\it{M$_{\odot}$}}). 
(8) Linewidth measured at 20\% of peak (km~s$^{-1}$); the symbol
    `*' denotes spectra with at least two kinematically resolved
    components, of which the $W20$ is for the main component (the one with
    the highest integrated flux).
(9) Dust mass multiplied by 100 (10$^9${\it{M$_{\odot}$}}). 
The 100$\times$\mdust\ of S+S pairs with only one detection equals to that of the detected component. 
The 100$\times$\mdust\ of S+E pairs includes only that of spiral component. 
(10) Stellar mass (10$^9${\it{M$_{\odot}$}}). 
The \mstar\ of S+E pairs includes only that of spiral component. 
(11) Star formation rate ($M_{\odot}$~yr$^{-1}$). 
The SFR of S+S pairs with only one detection equals to that of the detected component. 
The SFR of S+E pairs includes only that of spiral component. 
(12) Type of major-merger pairs. 
}
\end{center}
\end{table*}

\begin{table*}
\scriptsize
\begin{center}
\caption{Pairs from literatures.\label{tbl:lit_obs}}
\begin{tabular}{rrrrrrrrrrrrrrr}
\tableline\tableline
(1) & (2) & (3) & (4) & (5) & (6) & (7) & (8) & (9) & (10) & (11)   
\\
Pair ID & R.A. & decl. & 
$V_{\mathrm{H{\textsc i}}}$  & \mhi\  &  
100$\times$\mdust\  & 
 \mstar\   &    SFR    &     
 Type  & ref.  & beam size\\
 (H-KPAIR)  & (J2000)  & (J2000)  & (km\,s$^{-1}$)  & (10$^9${\it{M$_{\odot}$}})  & (10$^9${\it{M$_{\odot}$}})  & (10$^9${\it{M$_{\odot}$}})& ({\it{M$_{\odot}$}} yr$^{-1}$)  &  & 
   \\
\tableline
J0020+0049 & 00:20:27.4 & +00:49:59 & 5498 &  3.30$ \pm $0.49 &  1.90 &  31.24 &  1.39 & SE  & 1  & 3.3\arcmin \\
J0823+2120 & 08:23:32.6 & +21:20:16 & 5400 & 12.12$ \pm $1.17 &  3.96 &  33.79 &  3.98 & SS  & 2 & 10.0\arcmin \\
J0829+5531 & 08:29:15.0 & +55:31:21 & 7758 & 25.59$ \pm $2.65 & 21.48 &  85.47 &  3.98 & SS  & 1 &  10.0\arcmin \\
J1043+0645 & 10:43:51.9 & +06:46:00 & 8238 &  6.70$ \pm $0.27 & 12.34 &  59.98 &  5.01 & SS  & 3 &  3.5\arcmin$\times$ 3.8\arcmin \\
J1308+0422 & 13:08:28.3 & +04:22:01 & 7251 &  8.26$ \pm $0.26 &  4.93 &  23.58 &  0.66 & SS  & 3 &  3.5\arcmin$\times$ 3.8\arcmin\\
J1315+6207 & 13:15:34.6 & +62:07:28 & 9100 &  5.00$ \pm $0.83 & 12.77 & 103.90 & 62.89 & SS  & 2 &  10.0\arcmin \\
J1406+5043 & 14:06:21.7 & +50:43:29 & 1860 &  1.63$ \pm $0.17 &  2.09 &  13.92 &  0.92 & SE  & 1 &  10.0\arcmin \\
J1423+3400 & 14:23:42.5 & +34:00:30 & 3865 &  1.03$ \pm $0.10 &  2.90 &  25.92 &  1.23 & SS  & 1 &  3.3\arcmin \\
J1425+0313 & 14:25:05.5 & +03:13:59 & 10680 &  1.50$ \pm $0.35 &  $<1.02$ &  23.11 & $<0.10$ & SE  & 4 &  3.5\arcmin \\
J1444+1207 & 14:44:20.7 & +12:07:55 & 8895 &  5.93$ \pm $0.64 & 12.84 & 186.60 &  4.64 & SS  & 1 &  3.3\arcmin \\
J1608+2328 & 16:08:22.5 & +23:28:46 & 12121 & 14.87$ \pm $1.50 & 17.42 &  71.43 & 12.19 & SS  & 1 &  3.3\arcmin \\
J2047+0019 & 20:47:19.0 & +00:19:17 & 4204 & 22.44$ \pm $3.45 & 15.14 & 123.10 &  1.84 & SE  & 1 &  3.3\arcmin \\

\tableline
\end{tabular}
\tablecomments{Descriptions of Columns: 
(1) Pair ID. The designations are ``H-KPAIR J0020+0049", etc. 
(2) R.A. (h:m:s, J2000). 
(3) decl. (d:m:s, J2000). 
(4) \h1 velocity. 
(5) \h1 mass (10$^9${\it{M$_{\odot}$}}). 
(6) Dust mass multiplied by 100 (10$^9${\it{M$_{\odot}$}}). 
The 100$\times$\mdust\ of S+S pairs with only one detection equals to that of the detected component. 
The 100$\times$\mdust\ of S+E pairs includes only that of spiral component. 
(7) Stellar mass (10$^9${\it{M$_{\odot}$}}). The \mstar\ of S+E pairs includes only that of spiral component. 
(8) Star formation rate ($M_{\odot}$ yr$^{-1}$). 
The SFR of S+S pairs with only one detection equals to that of the detected component. 
The SFR of S+E pairs includes only that of spiral component. 
(9) Type of major-merger pairs. (10) References: 
1 \citet{Springob2005}; 
2  \citet{Huchtmeier1989}; 
3 \citet{2011Haynes}; 
4 \citet{Catinella2010}.  
}
\end{center}
\end{table*}

\clearpage

\begin{figure}[h]
\centering
\includegraphics[width = 0.55\textwidth]{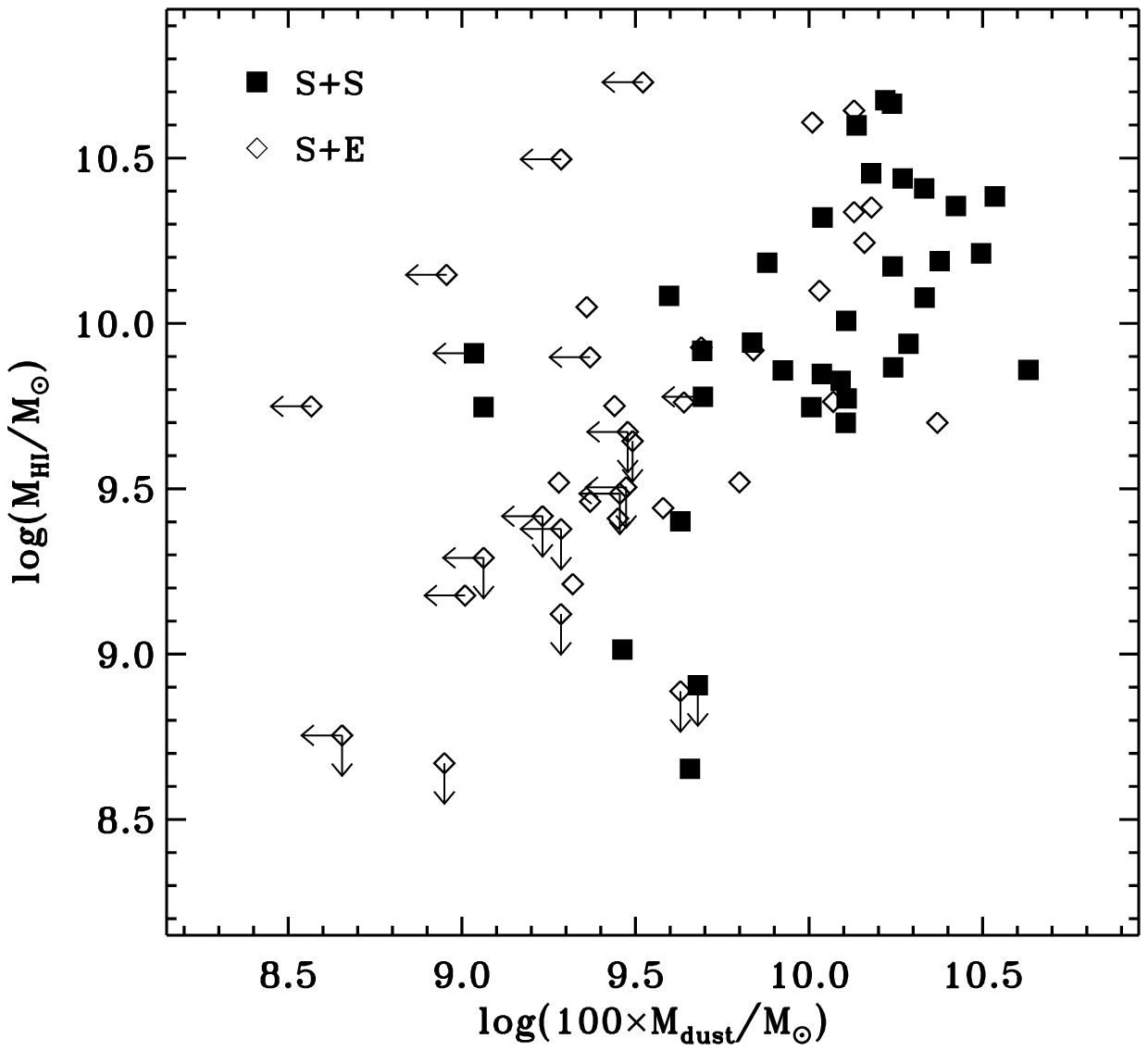}

\caption{
\h1 mass (\mhi) vs.\ $100\times$\mdust\ plot. 
}
\label{fig:mhi_mgas}
\end{figure}

\section{RESULTS}

Our final sample includes 70 pairs (34 S+S pairs, 36 S+E pairs)
whose \h1 mass \mhi\ can be found in Table~\ref{tbl:gbt_obs} and Table~\ref{tbl:lit_obs}. 
For S+S pairs,  
since the GBT beam cannot resolve them into individual galaxies, we treated each pair (including both components) as a single source. 
In these tables, we also listed  $100\times$\mdust\ as an estimate of total gas mass, 
stellar mass \mstar, and SFR, all taken from \citet{Cao2016}. 
The corresponding \mstar, $100\times$\mdust\ and SFR for each S+S pair are sums of the two components. 
In cases that one of the two ($100\times$\mdust\ or SFR) is undetected by 
{\it Herschel}, the true value for the pair should be limited 
between the detection and
the sum of the detection plus the upper-limit of the undetected component. 
A test has shown that, for our statistical results (Table~\ref{tbl:km_stats}), 
the difference between calculations adopting either of these two limits
is negligibly small (0.01-0.02 dex). We choose to take the detection as
the value of the pair. 
For each S+E pair, 
we assumed that the \h1\ mass is associated only with S component 
and contribution from the E component is negligible. 
We tested this assumption using the gas mass derived from the dust mass \citep{Cao2016}. 
We calculated the mean and error of the ratio of \mgas(E)/\mgas(S) 
using the K-M estimator \citep{Kaplan1958} which
exploits the information in the upperlimits of \mgas(E). 
The result is 0.11$\pm$0.01. This indicates that E galaxies 
contribute only 10\% of the gas mass of S+E pairs, which is indeed negligible. 
Other variables, including \mstar, $100\times$\mdust\ and SFR, are also for the S component only. 

In Table~\ref{tbl:km_stats}, 
means and errors of log[\mhi/($100\times$\mdust)], 
log(\mhi/\mstar), and log(\sfehi) are presented for the total sample of pairs and for the following three sets of contrasting sub-samples: 
(1) S+S vs.\ S+E; 
(2) log(\mstar/\msun) $<$ 10.7 vs.\ log(\mstar/\msun) $>$ 10.7 (for an S+S pair, \mstar\ is the mean of the two components); 
(3) ``JUS" vs.\ ``INT \& MER" pairs. 
According to \citet{Cao2016}, ``JUS" pairs are those without clear signs of interaction, 
``INT" and ``MER" are pairs with signs of interaction and merging. 
In order to exploit information in the upper limits, the calculations were carried out using the maximum likelihood Kaplan-Meier (K-M) estimator \citep{Kaplan1958}. 
We derived the survival curves of these ratios through VOStat \citep{VOStatDG2013}. 
The means and errors of statistical calculations are based on the integrated areas of the curves. 
The log[\mhi/($100\times$\mdust)] analysis is confined to sources (55) with
\mdust\ detections, and the log[SFR/\mhi] analysis to sources (58) with
\mhi\ detections. 

In Figure~\ref{fig:mhi_mgas} we compare \mhi\ (contamination corrected) with
$100\times$\mdust. There is a good correlation between the two
values, both probing the cold gas content in these pairs. The linear
correlation coefficient is 0.53. For the Spearman's rank correlation,
the coefficient is 0.61 and the significance is $2.60\times10^{-8}$.
The strong correlation suggests that a significant fraction of dust resides
in \h1\ gas, or the ratio between \mhi\ and \mh2\ is relatively
constant. For the total sample, the mean 
log[\mhi/($100\times$\mdust)]$=-0.06\pm 0.06$. \citet{Cao2016} adopted 
$100\times$\mdust\ as an estimate of the total gas mass \mgas.
Taken at face value, our result indicates that
the contribution of the \h1\ gas to the total gas mass is
$87(\pm 12)\%$. \citet{Draine2007} found an average
dust-to-gas mass ratio of 0.007 for nearby spiral galaxies, corresponding to
\mgas/\mdust\ $= 143$.  Assuming this gas-to-dust mass ratio for our
pairs, the contribution of the \h1\ gas to the total gas mass would be
$61(\pm 9)\%)$.

\begin{figure}[h]
\includegraphics[width = 0.48\textwidth]{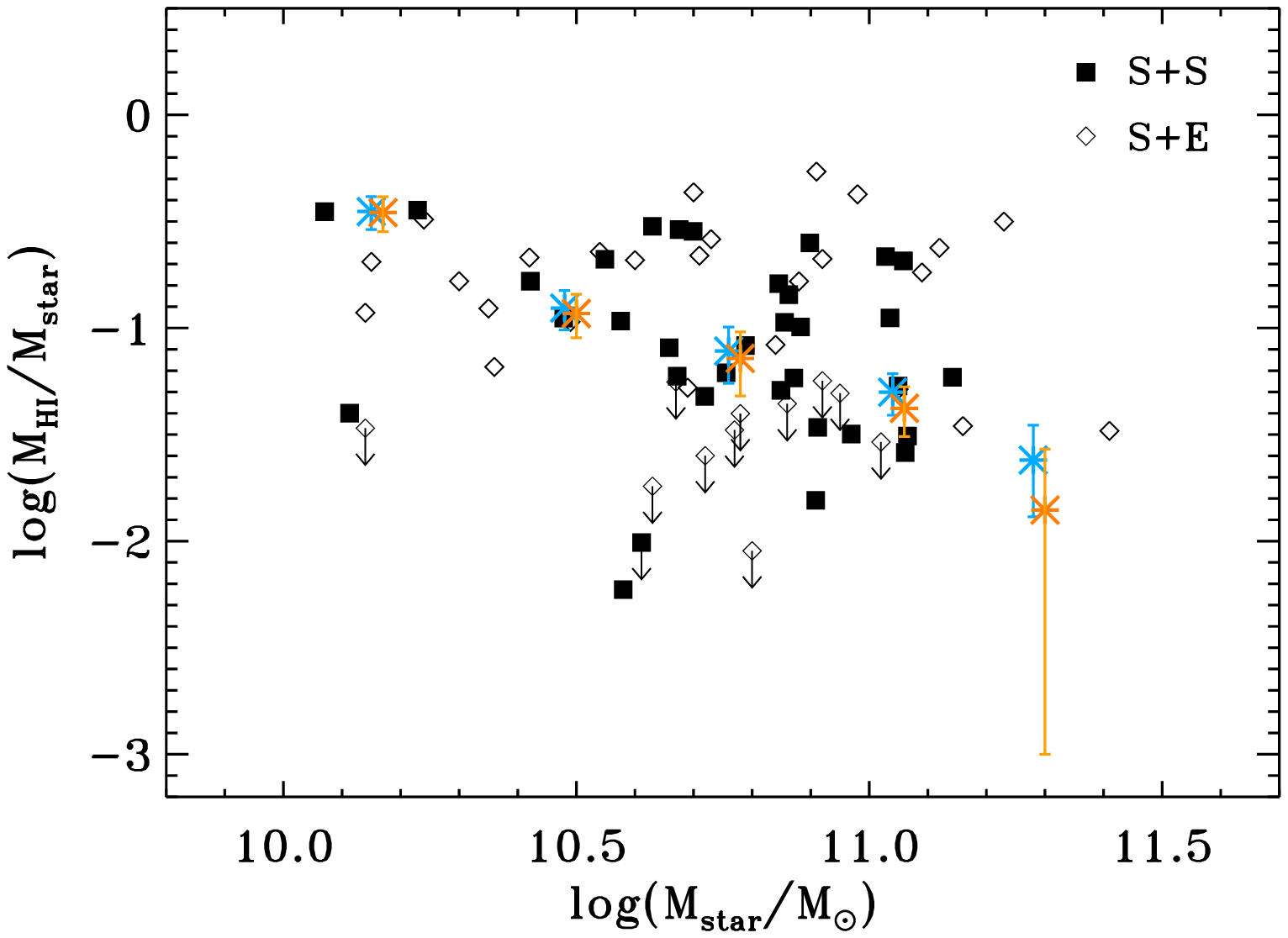}

\caption{Plot of log(\mhi/\mstar) vs. log(\mstar). 
For each S+E pair, \mstar\ includes only the 
stellar mass of the spiral component.
For each S+S pair, both the \h1 mass (\mhi) and the total
stellar mass (\mstar) of the pair are divided by 2. Results of \citet{Catinella2010}
are shown by light blue 8-point stars (means derived 
assuming fluxes of undetected sources
equal to the upperlimits) with error bars and by orange 8-point stars
(means derived assuming fluxes of undetected sources
equal to zero) with error bars.
}
\label{fig:hif_mstar}
\end{figure}

In Figure~\ref{fig:hif_mstar}, we plot log(\mhi/\mstar) against log(\mstar)
for S+S and S+E pairs using different symbols. Since each S+S pair
has two spirals while an S+E pair has only one, both \mhi\ and \mstar\ of
the former are divided by 2.  The mean log(\mhi/\mstar) ($=-1.12\pm
0.06$) for the total pair sample corresponds to a \h1\ gas fraction of
\fhi\ $= 7.6 (\pm 1.1)\%$. 
More massive paired galaxies (\mstar $>
10^{10.7}$\msun) have a slightly lower average (mean log(\mhi/\mstar)$=-1.13\pm
0.07$, corresponding to \fhi\ $= 7.4 (\pm 1.2)\%$) 
compared with less massive paired galaxies of \mstar $<
10^{10.7}$\msun\ which have mean log(\mhi/\mstar)$=-1.06\pm 0.11$
 (corresponding to \fhi\ $= 8.7 (\pm 2.2)\%$). 
This is consistent with the results of \citet{Catinella2010} who
showed that for a large sample of $\sim 1000$ SFGs of $10^{10} <$
\mstar $< 10^{11.5}$ \msun, there is a significant trend for the \h1
gas fraction to decrease with increasing \mstar, with an overall
average of \fhi\ $\sim 10\%$ (Figure~\ref{fig:hif_mstar}). 
We found no significant difference 
($<1\sigma$) between the means
of log(\mhi/\mstar) of S+S and of S+E pairs (Table~\ref{tbl:km_stats}).
This is different from \citet{Cao2016} who
found, although with only marginal significance, that
spirals in S+E pairs have on average lower total gas mass (estimated
using the dust mass) to stellar mass ratio than those in S+S pairs. 

\begin{figure}[h]
\includegraphics[width = 0.48\textwidth]{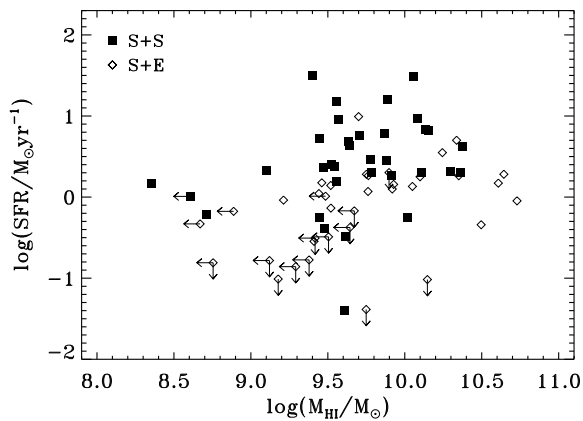}
\caption{Plot of log(SFR) vs.\ log(\mhi). 
For each S+E pair, SFR includes only that of the spiral component.
For each S+S pair, both the \h1\ mass (\mhi) and
the total SFR of the pair are divided by 2.}
\label{fig:mhi_sfr}
\end{figure}

In Figure~\ref{fig:mhi_sfr}, we present log(SFR) vs. log(\mhi) plot for S+S and S+E pairs. 
Here again, \mhi\ and \mstar\ of S+S pairs are divided by 2.
There is a sub-population of very active star-forming galaxies 
(with $\mathrm{SFR}\ \gtrsim 10 M_\odot\ \mathrm{yr^{-1}}$) in S+S pairs. 
These galaxies are largely missing in S+E pairs. 

\begin{figure}[h]
\includegraphics[width = 0.48\textwidth]{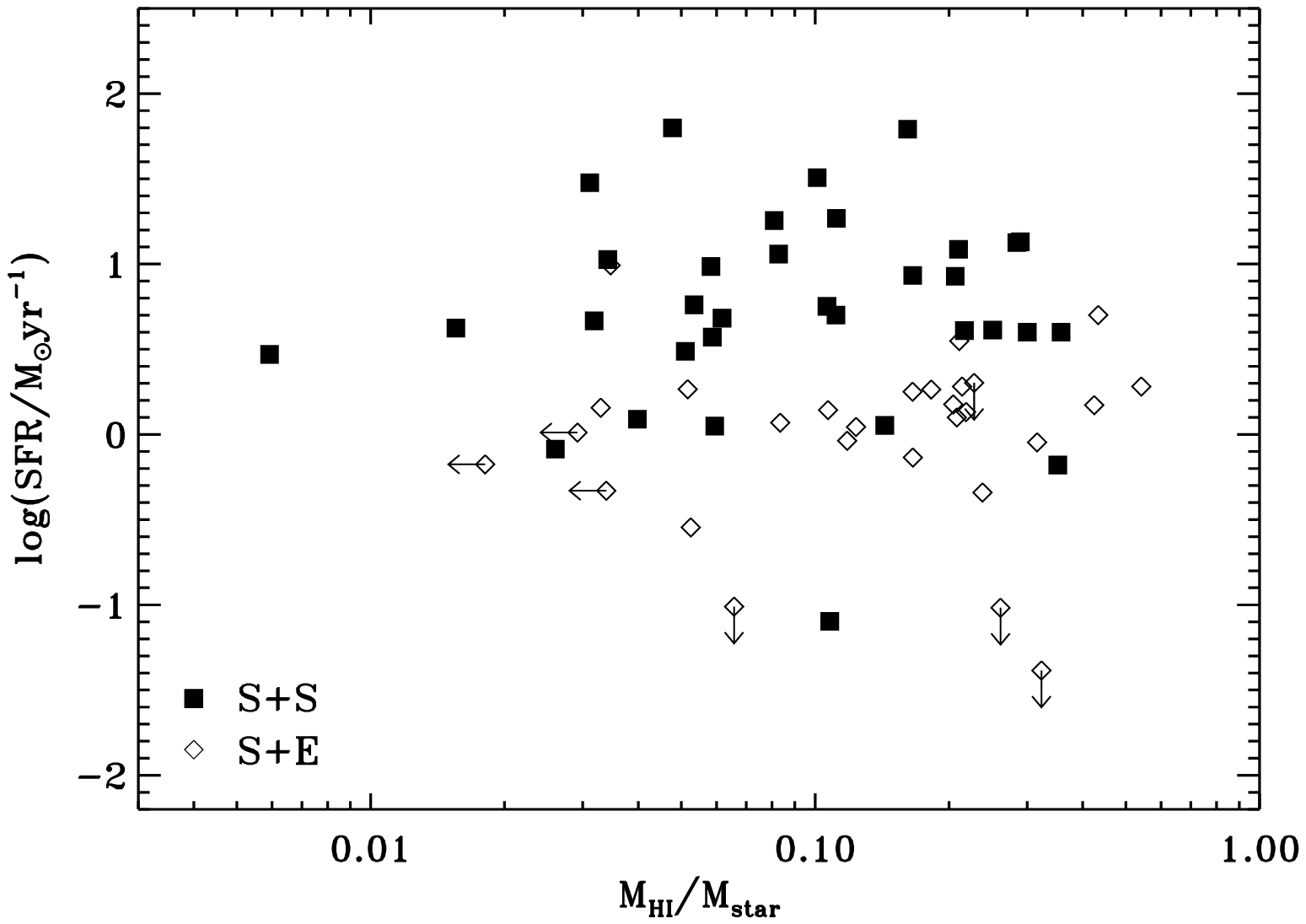}
\caption{Plot of log(SFR) vs.\ \mhi/\mstar. 
}
\label{fig:hif_sfr}
\end{figure}

In Figure~\ref{fig:hif_sfr}, log(SFR) is plotted against \h1 gas fraction \mhi/\mstar. 
It shows again that SFR of S+S pairs is 
systematically higher than that of S+E pairs. 
There is no clear dependence of SFR on \h1\ gas fraction. 

\begin{figure}[h]
\includegraphics[width = 0.48\textwidth]{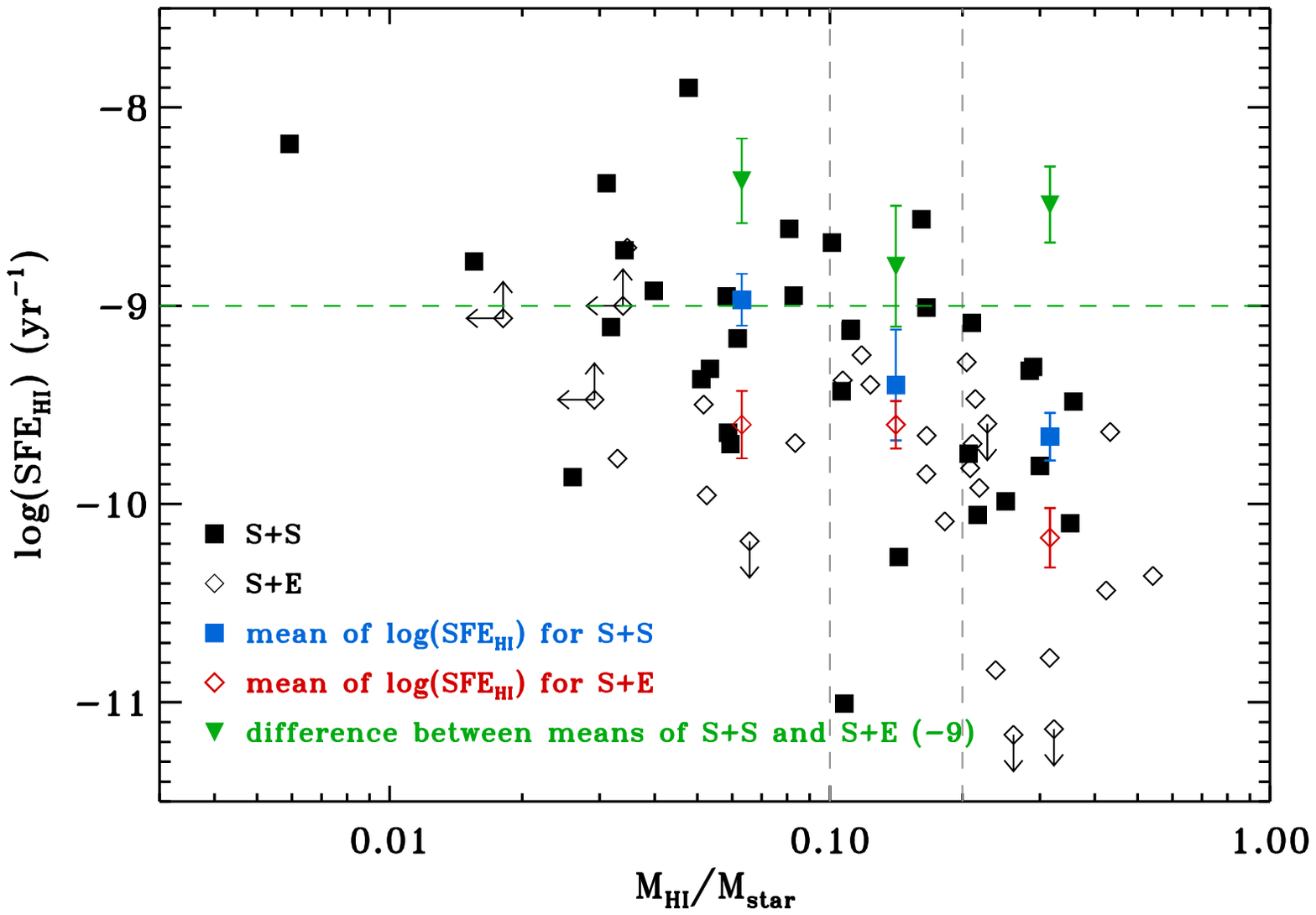}
\caption{Plot of log(\sfehi) vs. \mhi/\mstar. 
The blue filled squares with error bars represent means of 
log(\sfehi) of S+S pairs in \h1 fraction bins of \mhi/\mstar\ $<$~0.1, 0.1~$<$ \mhi/\mstar\ $<$~0.2, and \mhi/\mstar\ $>$ 0.2 for S+S pairs respectively. The red open diamonds with error bars are means for S+E pairs in the same bins. The vertical gray dashed lines at \mhi/\mstar$=$ 0.1 and 0.2 mark the boundaries between the 3 bins. The green filled downward triangles represent the differences between the means of S+S pairs and that of S+E pairs. The green dashed line at -9 of log(\sfehi) represents the zero level of the differences.
}
\label{fig:hif_sfehi}
\end{figure}

The ratio SFR/\mhi\ measures the star formation rate per unit \h1\ gas
mass, which will be referred as \sfehi\ hereafter.  In Figure~\ref{fig:hif_sfehi},
log(\sfehi) is plotted against the \h1 gas fraction \mhi/\mstar.  It
appears that for a given \mhi/\mstar, the
\sfehi\ of S+E pairs is systematically lower than that of S+S pairs,
and the difference is not sensitive to the \h1\ fraction.  As listed in
Table~\ref{tbl:km_stats}, the mean \sfehi\ of S+S pairs ($10^{-9.26\pm 0.11}\ 
\mathrm{yr}^{-1}$) is $\sim 4.6\times$ higher than that of S+E pairs
($10^{-9.92\pm 0.11}\ \mathrm{yr}^{-1}$). 
There is a small but insiginificant difference between the mean \sfehi\ of
pairs with signs of merger/interaction ($10^{-9.45\pm
  0.14}\ \mathrm{yr}^{-1}$) and that of pairs without
($10^{-9.60\pm 0.09}\ \mathrm{yr}^{-1}$). 
The mean \sfehi\ of the whole pair sample is $10^{-9.55\pm 0.09}\ \mathrm{yr}^{-1}$,
corresponding to a \h1\ consumption time of $3.5\pm0.7$~Gyrs.

\begin{table*}[t]
\scriptsize
\begin{center}

\caption{Kaplan-Meier estimation results. \label{tbl:km_stats}}
\begin{tabular}{crrrrrrrrrrrr}
\tableline\tableline
(1) & (2) & (3) & (4) & (5) & (6) & (7)  & (8)  & (9)  & (10)\\
Samples  & log[\mhi/(100$\times$\mdust)]  & Error  & Count  & log(\mhi/\mstar)  & Error  & Count  & log(SFR/\mhi)  & Error  & Count \\

\tableline
S+S                              &  -0.09  & 0.07  & 32  & -1.09  & 0.08  & 34  & -9.26  & 0.11  & 33  \\
S+E                              &   0.002  & 0.08  & 23  & -1.01 & 0.07  & 36  & -9.92  & 0.11 & 25    \\
\tableline
log\,\mstar $<$ 10.7  &  -0.12  & 0.10  & 23  & -1.06  & 0.11  & 28  & -9.59  & 0.14  & 24  \\
log\,\mstar $>$ 10.7  &  -0.02  & 0.06  & 32  & -1.13  & 0.07  & 42  & -9.53  & 0.12  & 34 \\
\tableline
JUS                              &  -0.05  & 0.09 & 28  &  -1.10 & 0.10  & 34  & -9.60  & 0.09  &  29 \\
INT \& MER                 & -0.07  & 0.07  & 27  & -1.11 & 0.07  & 36  & -9.45  & 0.14  &  29\\
\tableline 
Total                             & -0.06  & 0.06  & 55  & -1.12 & 0.06  & 70  & -9.55  & 0.09  & 58 \\
\tableline
\end{tabular}
\tablecomments{
Descriptions of Columns: 
(1) Samples. 
(2) Means of log[\mhi/(100$\times$\mdust)] (log[(100$\times$\mdust)/\mhi] for comparison sample). 
(5) Means of log(\mhi/\mstar).
(8) Means of log(SFR/\mhi), the units are yr$^{-1}$. 
(3), (6) and (9): Errors of the means.
(4), (7) and (10): Number of pairs in the calculation of the mean. 
}
\end{center}
\end{table*}

\section{DISCUSSION}

Given the large beam ($\rm FWHM = 9'$) of the GBT observations, the
\h1\ detected for each pair includes both gas inside the discs and the
stripped gas in tidal features and debris. 
Because of the exclusion of mergers with component separation less than 5~kpc and nearby pairs with recession velocity less than 2000~km s$^{-1}$, 
the H-KPAIR sample preferentially selects massive ($\gtrsim 10^{10}$\;\msun)
 early stage merger systems (before the final coalescence). 
Our observations show that these systems have similar \h1\ gas fractions
compared to normal spiral galaxies (Figure~\ref{fig:hif_mstar}).  For gas-rich dwarf galaxy pairs
\mstar\ $<10^{9.7}$\,\msun, \citet{Stierwalt2015} reached similar
results.  On the other hand, spiral galaxies in compact groups are
found to be \h1\ deficient \citep{Verdes-Montenegro2001,
  Borthakur2015, Walker2016}.

A major science goal of this study is to address the puzzling result
of the significant difference between the sSFR enhancement of spirals
in S+E pairs and in S+S pairs \citep{Xu2010, Cao2016}. Because all
pairs were selected using the same criteria regardless of
morphological type \citep{Domingue2009, Cao2016}, this difference cannot be due to any selection bias. 
One possibility is that the sSFR in
a paired galaxy is influenced by the immediate surrounding
environment.  This hypothesis is in agreement with the correlation
between sSFRs of the primaries and secondaries in major-merger S+S
pairs (i.e., the ``Holmberg effect" \citealt{Kennicutt1987,
  Xu2010}). On the other hand, \citet{Xu2010, Xu2012} did not find any
significant difference between the local densities around
S+E pairs and S+S pairs within projected radius of 2 Mpc. Therefore,
the linear scale of the environment effect must be less than 2 Mpc.
\citet{Xu2010, Xu2012} speculated that the IGM in the dark matter halo
(DMH) shared by both galaxies of a pair may play a significant role
here.  For example, when a DMH has strong (weak) cold flows 
\citep{Dekel2009, Keres2009}, galaxies inside it may have abundant
(scarce) cold gas supply to fuel active star formation. 
A prediction of this hypothesis is that spiral
galaxies in S+E pairs have systematically lower gas content than those
in S+S pairs. 

However, this is not supported by our result which shows no 
significant difference between the mean
log(\mhi/\mstar) of S+S and that of S+E pairs. 
It appears that the higher sSFR and \sfehi\ of S+S pairs are mainly 
due to a sub-population of very active SFGs, 
which are missing in S+E pairs (Figure~\ref{fig:mhi_sfr}). 
It will be very interesting to find
out how the high star formation enhancement in these pairs is
triggered, and why it is not happening in S+E pairs. 

Some insights can be gained from the examples studied by
\citet{Hibbard2001} using high resolution VLA \h1\ maps for galaxies
in the ``merger sequence". Three of the five systems in their sample
are early stage mergers with active star formation.  They all show
very extended tidal features in the \h1\ gas distribution.  However,
most active star formation is confined to the central region where
high density molecular gas and bright $\rm H_\alpha$ emission are
found.  Dynamical simulations of \citet{Olson1990} demonstrated that,
in the central $\sim$2~$\mathrm{kpc}$ of merging galaxies, interaction
induced collisions between gas clouds may play very important roles in
triggering enhanced star formation, and the effect is stronger in S+S
systems than in S+E systems.

\citet{Scudder2015} carried out VLA observations (beam $\rm
FWHM = 14\arcsec$) of the \h1\ 21~cm line emission for
34 galaxies in 17 nearby S+S pairs, and obtained 17 detections.
Compared to a control sample of galaxies, they found marginal evidence
(at $\sim$2$\sigma$ level) for a positive correlation between the \h1\
fraction and the SFR enhancement.  On the other hand, \citet{Cao2016}
did not see any significant correlation between SFR/\mgas\ enhancement
and gas fraction in H-KPAIR. 
Our results in Figure~\ref{fig:hif_sfr} (Figure~\ref{fig:hif_sfehi}) show 
also that the difference between SFR (\sfehi) of spirals in S+S and S+E pairs does not depend on \h1\ gas fraction.

\section{SUMMARY}

In this paper we present a study of the \h1\ gas content of a large K-band selected sample of galaxy pairs (H-KPAIRs). 
Among 88 pairs (44 S+S pairs, 44 S+E pairs), 
we observed 67 pairs using GBT for the 21~cm \h1\ fine-structure emission.
Except for 9 pairs that have severe RFIs and thus no informative signals could be extracted from the data,  
we derived \h1\ mass from the spectral line. 
The results include detections (46 pairs) and upper limits (12 pairs). 
In addition, \h1\ mass of other 12 pairs are collected from the literature. 
Compared with the $\it Herschel$ data of the same sample,
the relations between \mhi\ and $100\times$\mdust, \mhi\ and SFR, \mhi\ and \mstar\, and \sfehi\ and \h1\ fraction are studied. 
The means and errors of log[\mhi/$(100\times$\mdust)], log(\mhi/\mstar), and log(\sfehi) are derived and analyzed for the total sample and for three sets of contrasting sub-samples. 
The primary results are as follows.

\begin{description}
\item{1.} Both linear and Spearman rank correlation analyses show a significant 
correlation between \mhi\ and $100\times$\mdust. 
For the total sample, the mean 
log[\mhi/($100\times$\mdust)]$=-0.06\pm 0.06$, corresponding to a
\h1-to-total gas ratio of $87(\pm 12)\%)$ if 
the gas-to-dust mass ratio is assumed as 100. 

\item{2.} The mean \h1-to-stellar mass ratio of spirals in these pairs
  is $0.076\pm 0.011$, consistent with the average \h1\ gas fraction of
  spiral galaxies in general.  There is no significant
  difference ($<1\sigma$) between the means of log(\mhi/\mstar) of
  S+S and of S+E pairs (Table~\ref{tbl:km_stats}).

\item{3.} The mean \sfehi\ of S+S pairs ($10^{-9.26\pm 0.11}\,
\mathrm{yr}^{-1}$) is $\sim 4.6\times$ higher than that of S+E pairs
($10^{-9.92\pm 0.11}\ \mathrm{yr}^{-1}$), and the difference is not sensitive
to the \h1\ fraction.  A sub-population of
very active star-forming galaxies in S+S pairs are largely missing in
S+E pairs.

\item{4.} The difference between the mean \sfehi\ of pairs with signs of merger/interaction
  ($10^{-9.45\pm 0.14}\ \mathrm{yr}^{-1}$) and that of pairs without ($10^{-9.60\pm
    0.09}\ \mathrm{yr}^{-1}$) is insignificant ($< 1\sigma$). 

\item{5.} The mean \sfehi\ of the whole pair sample is $10^{-9.55\pm
    0.09}\ \mathrm{yr}^{-1}$, corresponding to a \h1\ consumption
  time of $3.5\pm0.7$ Gyrs.

\end{description}

\acknowledgments
This work is supported by National Key R\&D Program of China No.\ 2017YFA0402600, 
Open Project Program of the Key Laboratory of FAST, NAOC, Chinese Academy of Sciences, 
National Natural Science Foundation of China No. 11643003 and No.\ 11373038, and 
International Partnership Program of Chinese Academy of Sciences  
No.\ 114A11KYSB20160008. This work is sponsored in part 
by the Chinese Academy of Sciences (CAS), through a grant to
the CAS South America Center for Astronomy (CASSACA) in Santiago, Chile. 

UL acknowledges support by the research projects
AYA2014-53506-P from the Spanish Ministerio de Econom\'\i a y Competitividad,
from the European Regional Development Funds (FEDER)
and the Junta de Andaluc'\i a (Spain) grants FQM108.
D.~L.\ acknowledges support from "CAS Interdisciplinary Innovation Team" program. 
C.~C.\ is supported by NSFC-11503013, NSFC-11420101002.






\makeatletter 
\renewcommand{\thefigure}{A\@arabic\c@figure}
\makeatother

\appendix

\section{Appendix A}

In Figure~\ref{fig:appa_spectra_hi}, we present the full intensity spectra of 5 nearby normal galaxies NGC895, NGC2718, NGC3027, UGC10014, and NGC6140. 
\h1 observations and data reduction of these galaxies are the same as for the paired galaxies (Sections 3 \& 4). The results are compared with literature data in Table~\ref{tbl:appa_gbt_obs}. 
When there are more than one previous \h1 observation for a given galaxy, its literature data is chosen according the following order of preference: (1) the latest GBT observation, (2) the latest NRAO~91m observation, (3) the latest Arecibo observation, (4) the latest observation by other telescopes.
The comparison shows a systematic difference on a $\sim$15\% level, 
possibly due to a minor deviation in the calibration. 
This shall not significantly affect our main conclusions. 

\begin{figure*}[h]
\setcounter{figure}{0}

\includegraphics[width = 0.3333\textwidth]{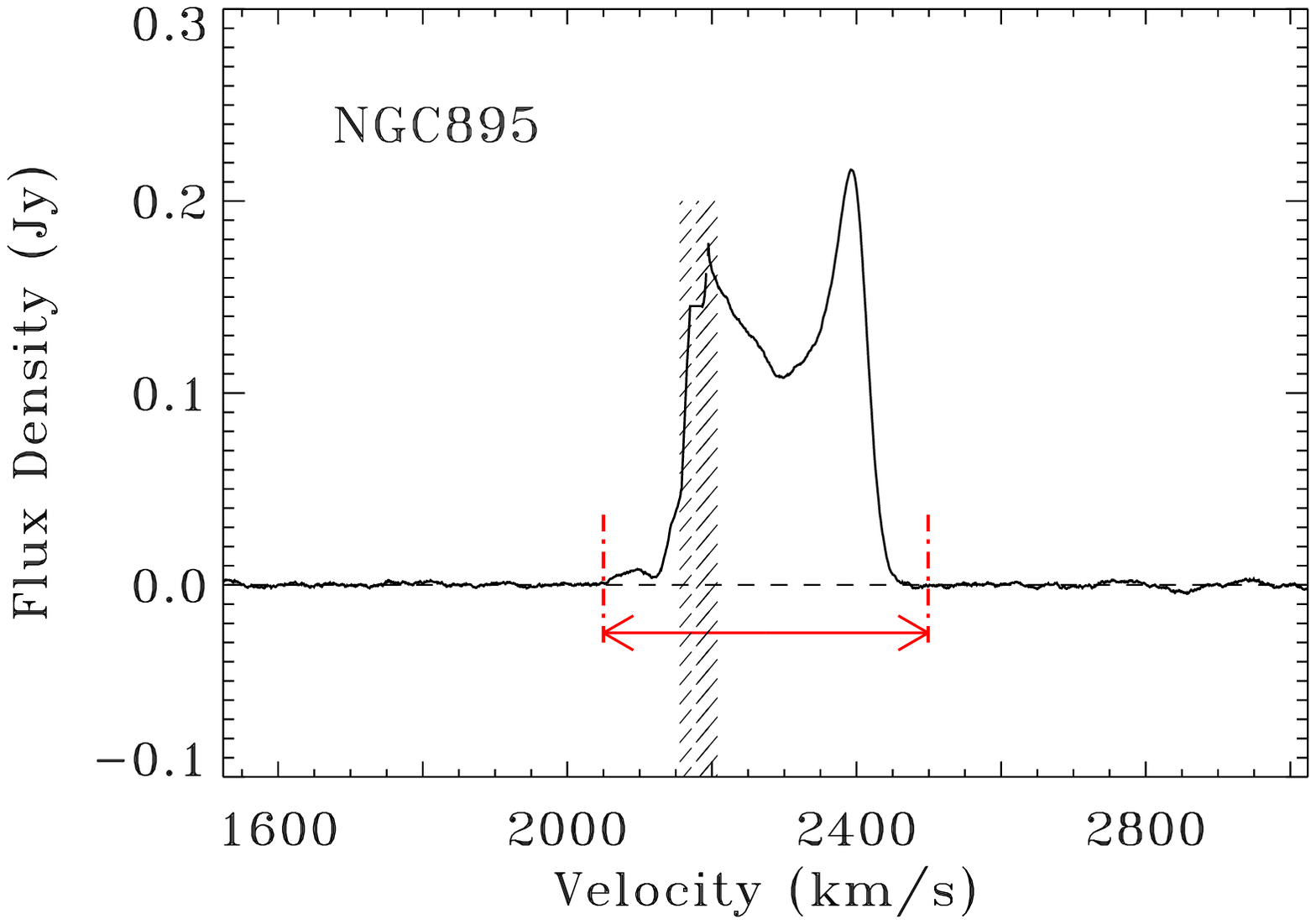}
\includegraphics[width = 0.3333\textwidth]{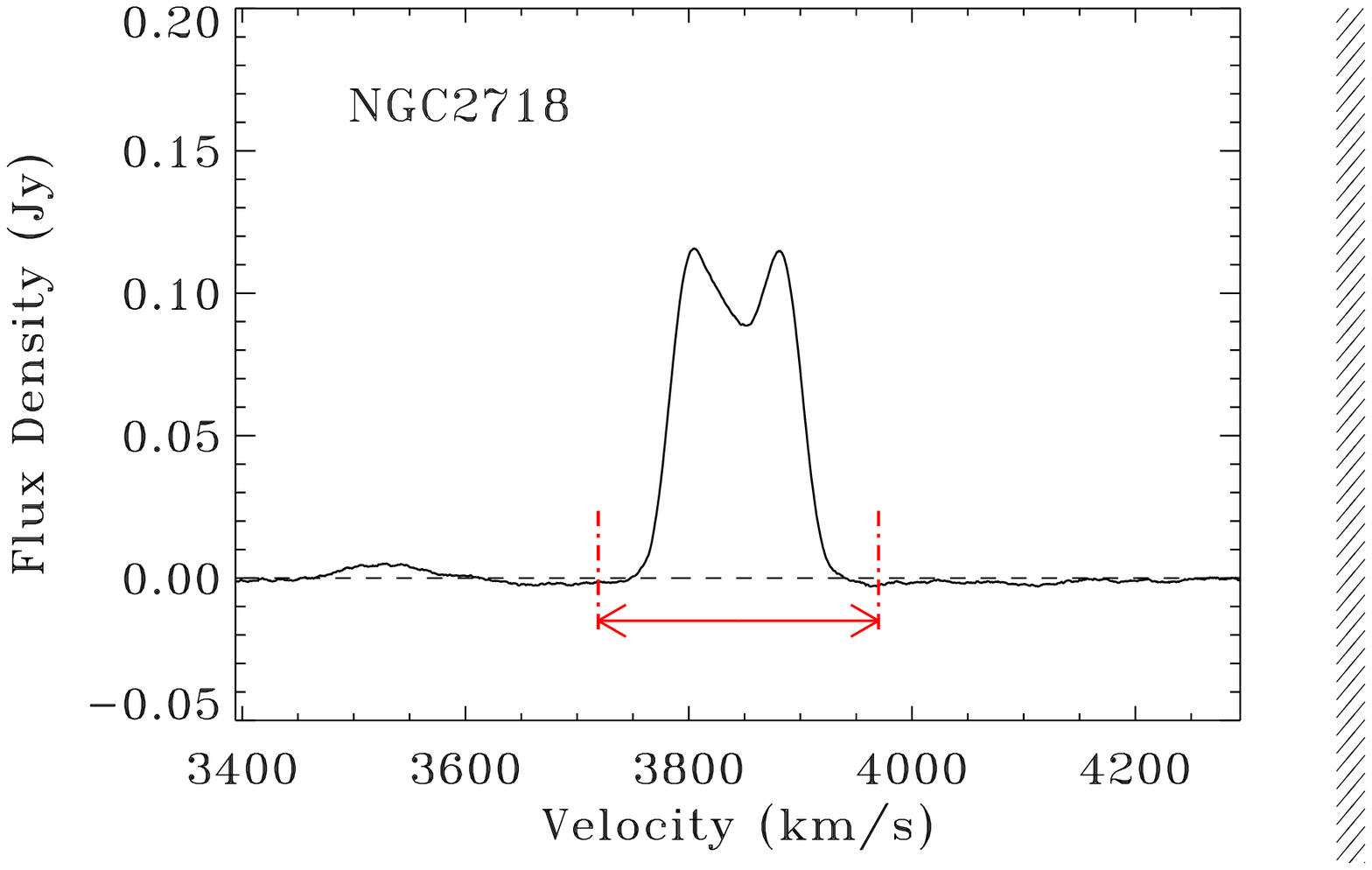}
\includegraphics[width = 0.3333\textwidth]{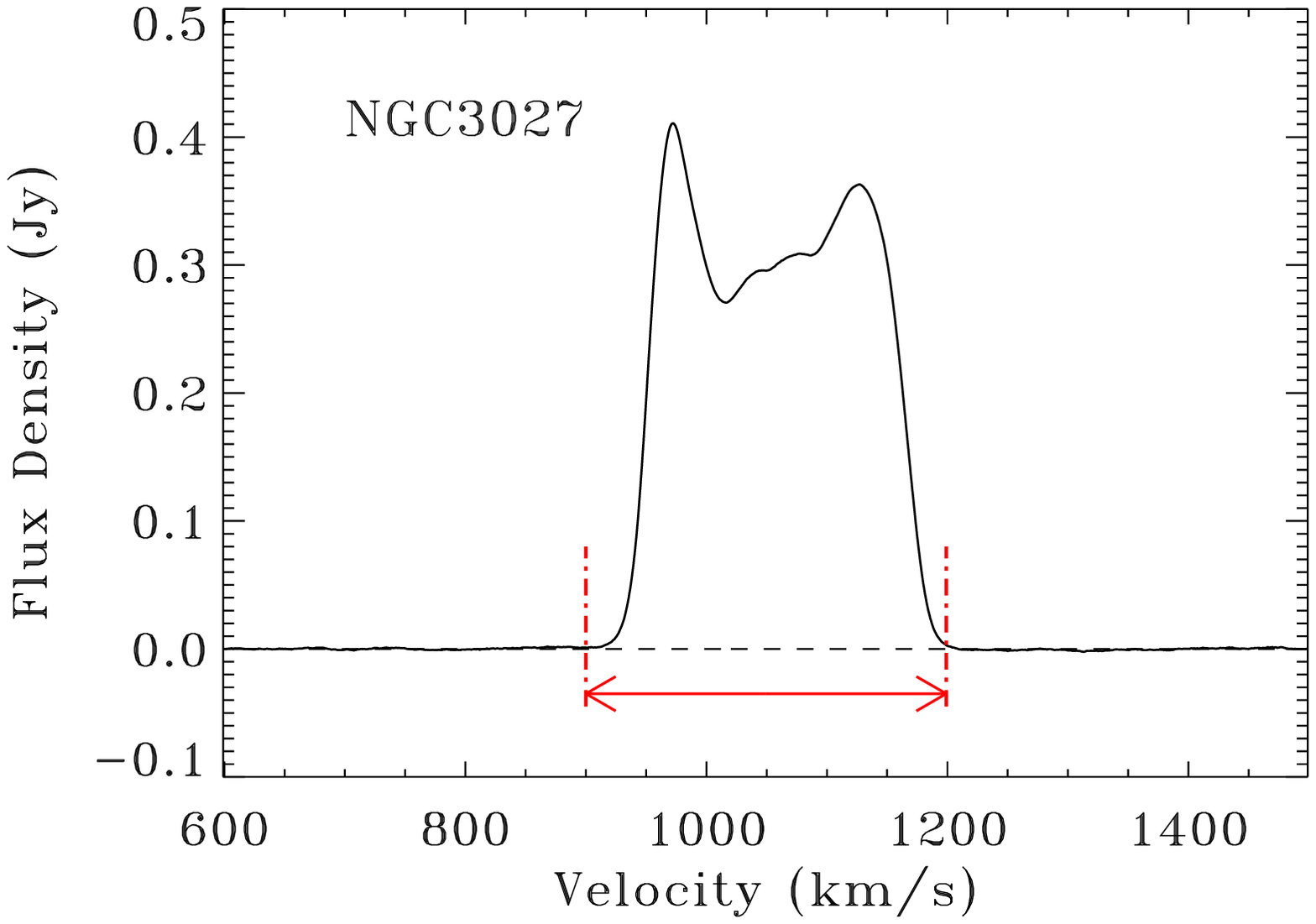}
\includegraphics[width = 0.3333\textwidth]{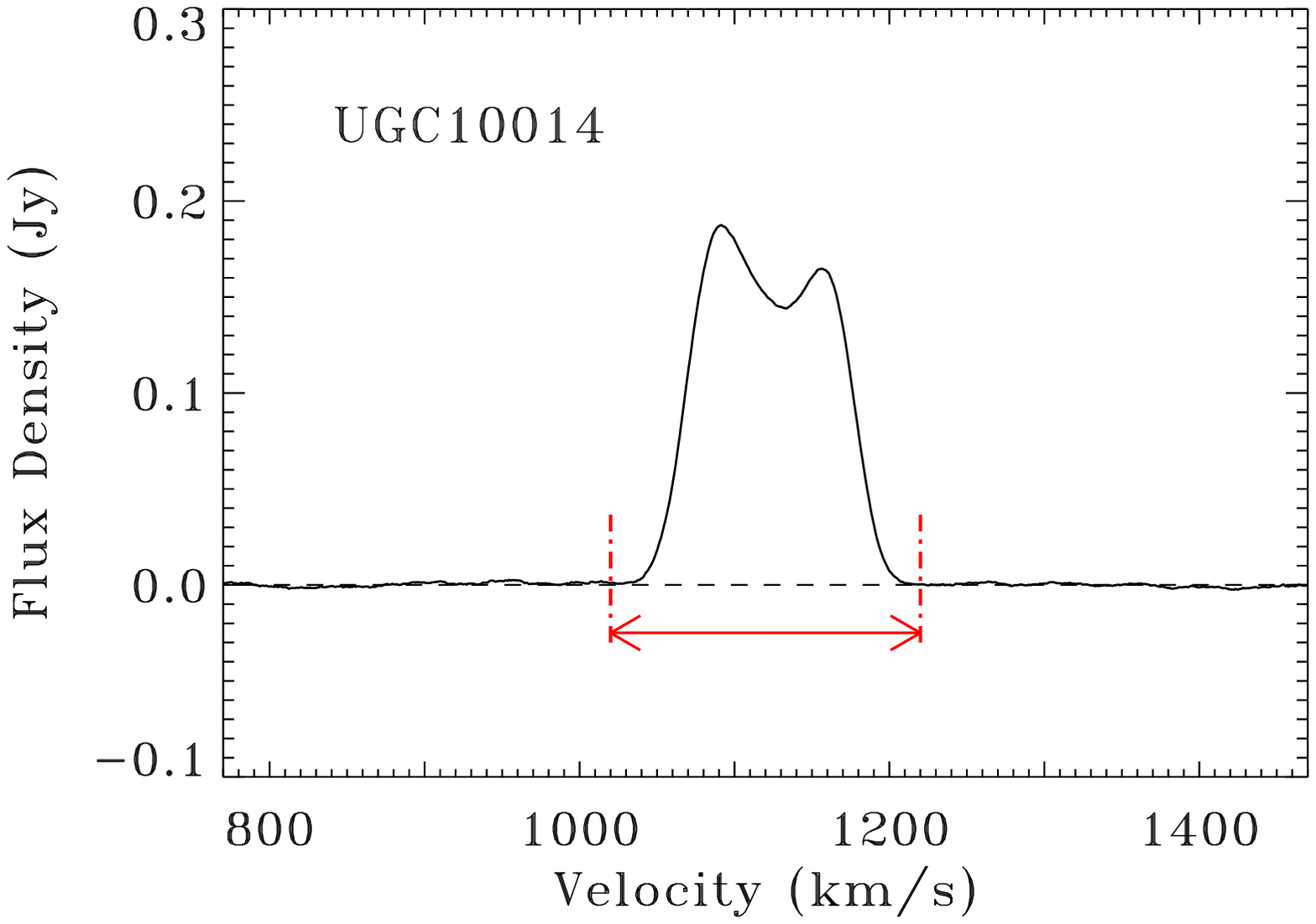}
\includegraphics[width = 0.3333\textwidth]{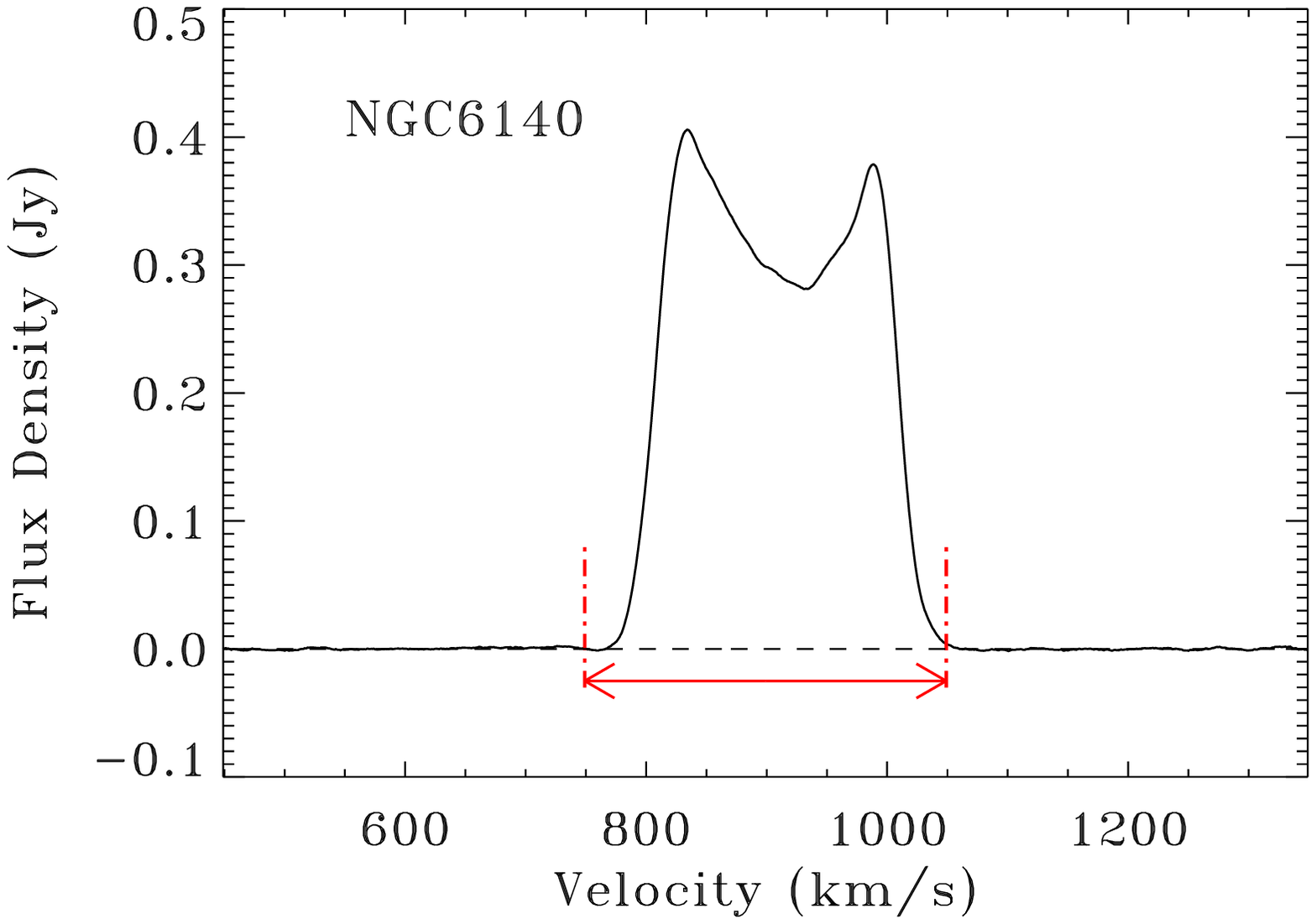}

\caption{\h1 profiles of the normal galaxies observed with the 100-m GBT. Regions with hash marks are frequencies where data are affected by RFI spikes. The Region between two red dot-dash lines represents the range of the intensity flux integration.
}
\label{fig:appa_spectra_hi}
\end{figure*}

\makeatletter 
\renewcommand{\thetable}{A\@arabic\c@figure}
\makeatother

\begin{table}
\tiny
\setcounter{table}{0}
\begin{center}
\caption{Nearby normal galaxies in GBT observations.\label{tbl:appa_gbt_obs}}
\begin{tabular}{crrrrrrrrrrrrrr}
\tableline\tableline
(1) & (2) & (3) & (4) & (5) & (6) & (7) & (8)  & (9) & (10)  & (11) \\
Galaxy ID & R.A. & decl. & $V_{\mathrm{optical}}$  & $V_{\mathrm{H{\textsc i}}}$  & 
d$_{j}$  & d$_{i}$  &
$Sdv_{\mathrm{H{\textsc i}}}$  & $Sdv_{\mathrm{H{\textsc i}, ref}}$  &
Dev.  & ref. \\ 
   & (J2000)  & (J2000)  & (km\,s$^{-1}$)  & (km\,s$^{-1}$)  & 
(\arcmin)  & (\arcmin)  & (Jy km\,s$^{-1}$)  & (Jy km\,s$^{-1}$)  &
   (\%)  \\ 

\tableline

NGC895  & 02:21:36.5  & -05:31:17  & 2288 & 2294  & 3.6  & 2.6  & 33.39  & 40.39  & 18.98  & GBT100m(1)  \\
NGC2718  & 08:58:50.5  & +06:17:35  & 3843 & 3842  & 1.74  & 0.91  & 13.07  & 15.67  & 18.09  & Arecibo(2)  \\
NGC3027 & 09:55:40.6  & +72:12:13  & 1058  & 1057  & 4.3  & 2.0  & 71.72  & 85.5  & 17.53  & NRAO91m(3) \\
UGC10014  & 15:45:43.9  & +12:30:38  & 1121  & 1121  & 1.3  & 1.2  & 19.00  & 21.2  & 10.95  & GB91m(4) \\
NGC6140  & 16:20:58.1  & +65:23:26  & 910  & 906  & 6.3  & 4.6  & 70.93  & 84.6  & 17.58  & NRAO91m(3)\\

\tableline
\end{tabular}
\tablecomments{Descriptions of Columns: (1) Galaxy ID. 
(2) R.A. (h:m:s, J2000). 
(3) decl. (d:m:s, J2000). 
(4) Optical velocity taken from SDSS or other telescopes. 
(5) \h1 mean velocity.
(6) The major diameter of the galaxy taken from NED. 
(7) The minor diameter of the galaxy taken from NED.
(8) Integrated \h1 line flux of our observation. 
(9) Integrated \h1  line flux of literature data.
(10) Relative difference between (8) and (9).
(11) References of the literature data: 1 \citet{2009Courtois}; 2 \citet{2011Haynes}; 3 \citet{1978Shostak}; 4 \citet{1988Tifft}.
}
\end{center}
\end{table}

\section{Appendix B}
\renewcommand\thefigure{\thesection\arabic{figure}}  

In this appendix, we present the algrorithm of the correction 
for contaminations due to neighboring galaxies, and the
postage stamp images (taken from SDSS-DR14) used in the
correction (Figure~\ref{fig:appb_sdss}). 

In order to correct for the contamination, we estimated the \h1 mass
of neighboring spiral galaxies using the following algorithm: First
the \h1-gas-to-stellar-mass ratio $G_{\mathrm{\h1}}/S$ is estimated by
log$_{10}(G_{\mathrm{\h1}}/S)
=-1.732\,38(g-r)+0.215\,182\mu_{i}-4.084\,51$ \citep{Zhang2009}, where
$\mu_{i}$ is the $i$-band surface brightness and $(g-r)$ is the
optical colour derived from the $g$- and $r$-band Petrosian
magnitudes.  The surface brightness used here is defined as
$\mu_i=m_i+2.5\mathrm{log}(2\pi R_{50}^2)$, where $m_i$ is the
apparent Petrosian $i$-band magnitude and $R_{50}$ the radius (in
units of arcsecond) enclosing 50 percent of the total Petrosian
$i$-band flux.  Then the stellar mass was estimated from the $i$-band
luminosity and $g-r$ colour using the formula $\mathrm{log}(M_*) =
\mathrm{log}(L_i)-0.222+0.864(g-r)$ \citep{Bell2003}.  The estimated
\h1\ mass was then multiplied by the GBT beam response function at the
distance of the galaxy, assuming the beam is a Gaussian with FWHM =
9\arcmin.  Finally, for each pair, the contamination due to \h1 mass
of neighboring spiral galaxies so estimated was subtracted from its
observed \h1\ mass. 

In Figure~\ref{fig:appb_sdss}, postage stamp images taken from SDSS-DR14 
are presented for individual pairs. In each image, paired galaxies are
marked with yellow letters and neighboring galaxies with redshifts 
inside the bandwidth of the \h1\ observation with green letters.
For pairs with \h1\ detections, the white circle represents the beam and
the red circle the searching circle ($r = 10$\arcmin). For pairs observed
but undetected by GBT, only the beam circle is plotted. For pairs with bad data,
neither circle is plotted. 
Besides, there are 9 pairs in galaxy groups and their \h1\ observations were not pointed to the pairs, which are excluded from our analysis. 

\renewcommand{\thefigure}{B1}

\figsetgrpstart
\figsetgrpnum{B1.1}
\figsetgrptitle{SDSS Image 1 of pairs}
\figsetplot{fB1_1.pdf}
\figsetgrpnote{Images collected by Pei Zuo.}
\figsetgrpend

\figsetend

\begin{figure*}

\figurenum{B1}
\includegraphics[width = 0.75\textwidth]{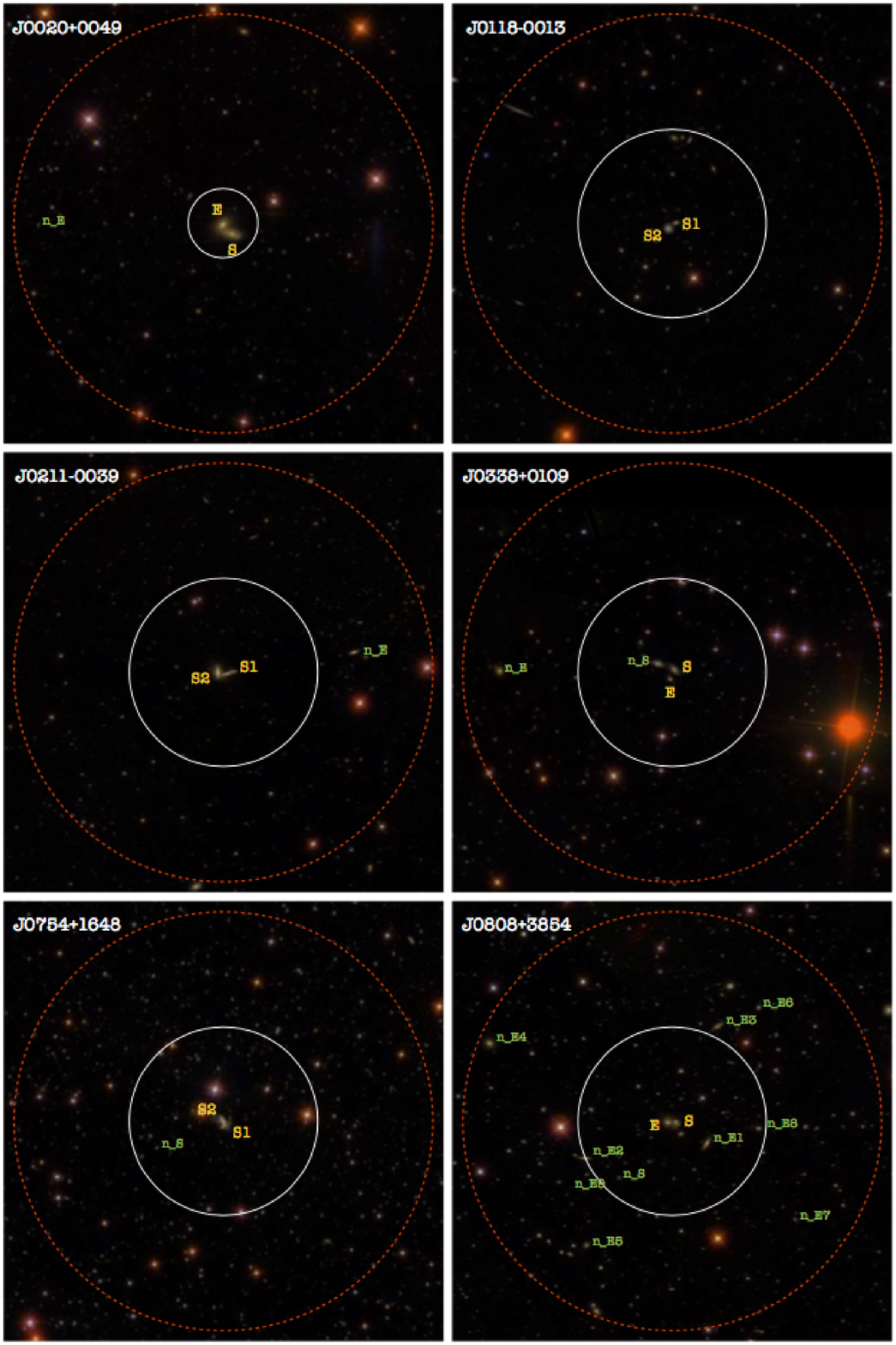}
\centering
\caption{SDSS images of the 88 pairs. 
The white circles represent the FWHMs of the beams for the telescopes, 
which include GBT and the data from the literature (except for the bad data). 
The center of these circles are at the pointing position of the data. 
The red dashed line circles represent the search radius of 10\arcmin\ for the nearby galaxies. 
For S+E pairs, ``S" and ``E" are for the S and E components, respectively. 
For S+S pairs, ``S1" represents the western galaxy and ``S2" represents the eastern one. 
The nearby spirals and ellipticals are denoted by green n\_S1, n\_E1, n\_S2, n\_E2, ..., 
and only neighbors with redshift falling into the GBT bandpass. 
(The complete figure set (15 images) is available.)}
\label{fig:appb_sdss}
\end{figure*}


\begin{thebibliography}{}
\expandafter\ifx\csname natexlab\endcsname\relax\def\natexlab#1{#1}\fi

\bibitem[{{Alonso} {et~al.}(2004){Alonso}, {Tissera}, {Coldwell}, \&
  {Lambas}}]{Alonso2004}
{Alonso}, M.~S., {Tissera}, P.~B., {Coldwell}, G., \& {Lambas}, D.~G. 2004,
  \mnras, 352, 1081

\bibitem[{{Barton} {et~al.}(2000){Barton}, {Geller}, \& {Kenyon}}]{Barton2000}
{Barton}, E.~J., {Geller}, M.~J., \& {Kenyon}, S.~J. 2000, \apj, 530, 660

\bibitem[{{Bell} {et~al.}(2003){Bell}, {McIntosh}, {Katz}, \&
  {Weinberg}}]{Bell2003}
{Bell}, E.~F., {McIntosh}, D.~H., {Katz}, N., \& {Weinberg}, M.~D. 2003, \apjs,
  149, 289

\bibitem[{{Bergvall} {et~al.}(2003){Bergvall}, {Laurikainen}, \&
  {Aalto}}]{Bergvall2003}
{Bergvall}, N., {Laurikainen}, E., \& {Aalto}, S. 2003, \aap, 405, 31

\bibitem[{{Borthakur} {et~al.}(2015){Borthakur}, {Yun}, {Verdes-Montenegro},
  {Heckman}, {Zhu}, \& {Braatz}}]{Borthakur2015}
{Borthakur}, S., {Yun}, M.~S., {Verdes-Montenegro}, L., {et~al.} 2015, \apj,
  812, 78

\bibitem[{{Bushouse} {et~al.}(1988){Bushouse}, {Werner}, \&
  {Lamb}}]{Bushouse1988}
{Bushouse}, H.~A., {Werner}, M.~W., \& {Lamb}, S.~A. 1988, \apj, 335, 74

\bibitem[{{Cao} {et~al.}(2016){Cao}, {Xu}, {Domingue}, {Buat}, {Cheng}, {Gao},
  {Huang}, {Jarrett}, {Lisenfeld}, {Lu}, {Mazzarella}, {Sun}, {Wu}, {Yun},
  {Ronca}, \& {Jacques}}]{Cao2016}
{Cao}, C., {Xu}, C.~K., {Domingue}, D., {et~al.} 2016, \apjs, 222, 16

\bibitem[{{Catinella} {et~al.}(2010){Catinella}, {Schiminovich}, {Kauffmann},
  {Fabello}, {Wang}, {Hummels}, {Lemonias}, {Moran}, {Wu}, {Giovanelli},
  {Haynes}, {Heckman}, {Basu-Zych}, {Blanton}, {Brinchmann}, {Budav{\'a}ri},
  {Gon{\c c}alves}, {Johnson}, {Kennicutt}, {Madore}, {Martin}, {Rich},
  {Tacconi}, {Thilker}, {Wild}, \& {Wyder}}]{Catinella2010}
{Catinella}, B., {Schiminovich}, D., {Kauffmann}, G., {et~al.} 2010, \mnras,
  403, 683

\bibitem[{{Condon} \& {Ransom}(2016)}]{Condon2016}
{Condon}, J.~J., \& {Ransom}, S.~M. 2016, {Essential Radio Astronomy}

\bibitem[{{Courtois} {et~al.}(2009){Courtois}, {Tully}, {Fisher}, {Bonhomme},
  {Zavodny}, \& {Barnes}}]{2009Courtois}
{Courtois}, H.~M., {Tully}, R.~B., {Fisher}, J.~R., {et~al.} 2009, \aj, 138,
  1938

\bibitem[{{Dasyra} {et~al.}(2006){Dasyra}, {Tacconi}, {Davies}, {Genzel},
  {Lutz}, {Naab}, {Burkert}, {Veilleux}, \& {Sanders}}]{Dasyra2006}
{Dasyra}, K.~M., {Tacconi}, L.~J., {Davies}, R.~I., {et~al.} 2006, \apj, 638,
  745

\bibitem[{{Dekel} {et~al.}(2009){Dekel}, {Birnboim}, {Engel}, {Freundlich},
  {Goerdt}, {Mumcuoglu}, {Neistein}, {Pichon}, {Teyssier}, \&
  {Zinger}}]{Dekel2009}
{Dekel}, A., {Birnboim}, Y., {Engel}, G., {et~al.} 2009, \nat, 457, 451

\bibitem[{{Domingue} {et~al.}(2016){Domingue}, {Cao}, \& {Xu}}]{Domingue2016}
{Domingue}, D.~L., {Cao}, C., \& {Xu}, C.~K. e.~a. 2016, \apj, 829, 78

\bibitem[{{Domingue} {et~al.}(2009){Domingue}, {Xu}, {Jarrett}, \&
  {Cheng}}]{Domingue2009}
{Domingue}, D.~L., {Xu}, C.~K., {Jarrett}, T.~H., \& {Cheng}, Y. 2009, \apj,
  695, 1559

\bibitem[{{Draine} {et~al.}(2007){Draine}, {Dale}, {Bendo}, {Gordon}, {Smith},
  {Armus}, {Engelbracht}, {Helou}, {Kennicutt}, {Li}, {Roussel}, {Walter},
  {Calzetti}, {Moustakas}, {Murphy}, {Rieke}, {Bot}, {Hollenbach}, {Sheth}, \&
  {Teplitz}}]{Draine2007}
{Draine}, B.~T., {Dale}, D.~A., {Bendo}, G., {et~al.} 2007, \apj, 663, 866

\bibitem[{{Ellison} {et~al.}(2008){Ellison}, {Patton}, {Simard}, \&
  {McConnachie}}]{Ellison2008}
{Ellison}, S.~L., {Patton}, D.~R., {Simard}, L., \& {McConnachie}, A.~W. 2008,
  \aj, 135, 1877

\bibitem[{{Ellison} {et~al.}(2010){Ellison}, {Patton}, {Simard}, {McConnachie},
  {Baldry}, \& {Mendel}}]{Ellison2010}
{Ellison}, S.~L., {Patton}, D.~R., {Simard}, L., {et~al.} 2010, \mnras, 407,
  1514

\bibitem[{{Haynes} \& {Herter}(1988)}]{Haynes1988}
{Haynes}, M.~P., \& {Herter}, T. 1988, \aj, 96, 504

\bibitem[{{Haynes} {et~al.}(2011){Haynes}, {Giovanelli}, {Martin}, {Hess},
  {Saintonge}, {Adams}, {Hallenbeck}, {Hoffman}, {Huang}, {Kent}, {Koopmann},
  {Papastergis}, {Stierwalt}, {Balonek}, {Craig}, {Higdon}, {Kornreich},
  {Miller}, {O'Donoghue}, {Olowin}, {Rosenberg}, {Spekkens}, {Troischt}, \&
  {Wilcots}}]{2011Haynes}
{Haynes}, M.~P., {Giovanelli}, R., {Martin}, A.~M., {et~al.} 2011, \aj, 142,
  170

\bibitem[{{Hibbard} {et~al.}(2001){Hibbard}, {van der Hulst}, {Barnes}, \&
  {Rich}}]{Hibbard2001}
{Hibbard}, J.~E., {van der Hulst}, J.~M., {Barnes}, J.~E., \& {Rich}, R.~M.
  2001, \aj, 122, 2969

\bibitem[{{Huchtmeier} \& {Richter}(1989)}]{Huchtmeier1989}
{Huchtmeier}, W.~K., \& {Richter}, O.-G. 1989, {A General Catalog of HI
  Observations of Galaxies. The Reference Catalog.}, 350

\bibitem[{{Hummel}(1981)}]{Hummel1981}
{Hummel}, E. 1981, \aap, 96, 111

\bibitem[{Kaplan \& Meier(1958)}]{Kaplan1958}
Kaplan, E.~L., \& Meier, P. 1958, Journal of the American Statistical
  Association, 53, 457

\bibitem[{{Keel} {et~al.}(1985){Keel}, {Kennicutt}, {Hummel}, \& {van der
  Hulst}}]{Keel1985}
{Keel}, W.~C., {Kennicutt}, Jr., R.~C., {Hummel}, E., \& {van der Hulst}, J.~M.
  1985, \aj, 90, 708

\bibitem[{{Kennicutt} {et~al.}(1987){Kennicutt}, {Roettiger}, {Keel}, {van der
  Hulst}, \& {Hummel}}]{Kennicutt1987}
{Kennicutt}, Jr., R.~C., {Roettiger}, K.~A., {Keel}, W.~C., {van der Hulst},
  J.~M., \& {Hummel}, E. 1987, \aj, 93, 1011

\bibitem[{{Kere{\v s}} {et~al.}(2009){Kere{\v s}}, {Katz}, {Fardal},
  {Dav{\'e}}, \& {Weinberg}}]{Keres2009}
{Kere{\v s}}, D., {Katz}, N., {Fardal}, M., {Dav{\'e}}, R., \& {Weinberg},
  D.~H. 2009, \mnras, 395, 160

\bibitem[{{Lambas} {et~al.}(2003){Lambas}, {Tissera}, {Alonso}, \&
  {Coldwell}}]{Lambas2003}
{Lambas}, D.~G., {Tissera}, P.~B., {Alonso}, M.~S., \& {Coldwell}, G. 2003,
  \mnras, 346, 1189

\bibitem[{{Larson} \& {Tinsley}(1978)}]{Larson1978}
{Larson}, R.~B., \& {Tinsley}, B.~M. 1978, \apj, 219, 46

\bibitem[{{Li} {et~al.}(2008){Li}, {Kauffmann}, {Heckman}, {Jing}, \&
  {White}}]{Li2008}
{Li}, C., {Kauffmann}, G., {Heckman}, T.~M., {Jing}, Y.~P., \& {White},
  S.~D.~M. 2008, \mnras, 385, 1903

\bibitem[{{Marganian} {et~al.}(2006){Marganian}, {Garwood}, {Braatz},
  {Radziwill}, \& {Maddalena}}]{Marganian2006}
{Marganian}, P., {Garwood}, R.~W., {Braatz}, J.~A., {Radziwill}, N.~M., \&
  {Maddalena}, R.~J. 2006, in Astronomical Society of the Pacific Conference
  Series, Vol. 351, Astronomical Data Analysis Software and Systems XV, ed.
  C.~{Gabriel}, C.~{Arviset}, D.~{Ponz}, \& S.~{Enrique}, 512

\bibitem[{{Nikolic} {et~al.}(2004){Nikolic}, {Cullen}, \&
  {Alexander}}]{Nikolic2004}
{Nikolic}, B., {Cullen}, H., \& {Alexander}, P. 2004, \mnras, 355, 874

\bibitem[{{Olson} \& {Kwan}(1990)}]{Olson1990}
{Olson}, K.~M., \& {Kwan}, J. 1990, \apj, 361, 426

\bibitem[{{Patton} {et~al.}(2013){Patton}, {Torrey}, {Ellison}, {Mendel}, \&
  {Scudder}}]{Patton2013}
{Patton}, D.~R., {Torrey}, P., {Ellison}, S.~L., {Mendel}, J.~T., \& {Scudder},
  J.~M. 2013, \mnras, 433, L59

\bibitem[{{Sanders} \& {Mirabel}(1996)}]{Sanders1996}
{Sanders}, D.~B., \& {Mirabel}, I.~F. 1996, \araa, 34, 749

\bibitem[{{Scudder} {et~al.}(2015){Scudder}, {Ellison}, {Momjian}, {Rosenberg},
  {Torrey}, {Patton}, {Fertig}, \& {Mendel}}]{Scudder2015}
{Scudder}, J.~M., {Ellison}, S.~L., {Momjian}, E., {et~al.} 2015, \mnras, 449,
  3719

\bibitem[{{Scudder} {et~al.}(2012){Scudder}, {Ellison}, {Torrey}, {Patton}, \&
  {Mendel}}]{Scudder2012}
{Scudder}, J.~M., {Ellison}, S.~L., {Torrey}, P., {Patton}, D.~R., \& {Mendel},
  J.~T. 2012, \mnras, 426, 549

\bibitem[{{Shostak}(1978)}]{1978Shostak}
{Shostak}, G.~S. 1978, \aap, 68, 321

\bibitem[{{Springob} {et~al.}(2005){Springob}, {Haynes}, {Giovanelli}, \&
  {Kent}}]{Springob2005}
{Springob}, C.~M., {Haynes}, M.~P., {Giovanelli}, R., \& {Kent}, B.~R. 2005,
  VizieR Online Data Catalog, 8077

\bibitem[{{Stierwalt} {et~al.}(2015){Stierwalt}, {Besla}, {Patton}, {Johnson},
  {Kallivayalil}, {Putman}, {Privon}, \& {Ross}}]{Stierwalt2015}
{Stierwalt}, S., {Besla}, G., {Patton}, D., {et~al.} 2015, \apj, 805, 2

\bibitem[{{Sulentic}(1989)}]{Sulentic1989}
{Sulentic}, J.~W. 1989, \aj, 98, 2066

\bibitem[{{Telesco} {et~al.}(1988){Telesco}, {Wolstencroft}, \&
  {Done}}]{Telesco1988}
{Telesco}, C.~M., {Wolstencroft}, R.~D., \& {Done}, C. 1988, \apj, 329, 174

\bibitem[{{Tifft} \& {Cocke}(1988)}]{1988Tifft}
{Tifft}, W.~G., \& {Cocke}, W.~J. 1988, \apjs, 67, 1

\bibitem[{{Verdes-Montenegro} {et~al.}(2001){Verdes-Montenegro}, {Yun},
  {Williams}, {Huchtmeier}, {Del Olmo}, \& {Perea}}]{Verdes-Montenegro2001}
{Verdes-Montenegro}, L., {Yun}, M.~S., {Williams}, B.~A., {et~al.} 2001, \aap,
  377, 812

\bibitem[{{VOStat Development Group}(2013)}]{VOStatDG2013}
{VOStat Development Group}. 2013, {VOStat: Statistical analysis of astronomical
  data}, Astrophysics Source Code Library, ascl:1309.008

\bibitem[{{Walker} {et~al.}(2016){Walker}, {Johnson}, {Gallagher}, {Privon},
  {Kepley}, {Whelan}, {Desjardins}, \& {Zabludoff}}]{Walker2016}
{Walker}, L.~M., {Johnson}, K.~E., {Gallagher}, S.~C., {et~al.} 2016, \aj, 151,
  30

\bibitem[{{Xu} \& {Sulentic}(1991)}]{Xu1991}
{Xu}, C., \& {Sulentic}, J.~W. 1991, \apj, 374, 407

\bibitem[{{Xu} {et~al.}(2012){Xu}, {Zhao}, {Scoville}, {Capak}, {Drory}, \&
  {Gao}}]{Xu2012}
{Xu}, C.~K., {Zhao}, Y., {Scoville}, N., {et~al.} 2012, \apj, 747, 85

\bibitem[{{Xu} {et~al.}(2010){Xu}, {Domingue}, {Cheng}, {Lu}, {Huang}, {Gao},
  {Mazzarella}, {Cutri}, {Sun}, \& {Surace}}]{Xu2010}
{Xu}, C.~K., {Domingue}, D., {Cheng}, Y.-W., {et~al.} 2010, \apj, 713, 330

\bibitem[{{Zhang} {et~al.}(2009){Zhang}, {Li}, {Kauffmann}, {Zou}, {Catinella},
  {Shen}, {Guo}, \& {Chang}}]{Zhang2009}
{Zhang}, W., {Li}, C., {Kauffmann}, G., {et~al.} 2009, \mnras, 397, 1243

\end{thebibliography}
\end{document}